\documentclass{aa}          %%aaaaaaaaaaaaaaaaaaaaaaaaaaaaaaaaaaaaa(to publish)
\usepackage{tabularx}
\usepackage{graphicx}
\usepackage{natbib}
\usepackage{txfonts}
\usepackage{longtable}
\usepackage{multirow}
\usepackage{lscape}
\usepackage{amssymb}
\usepackage{bm}
\usepackage{color}
\renewcommand*{\figurename}{ Fig.}
\renewcommand*{\tablename}{Table}
\newlength{\pointwidth}
\DeclareGraphicsExtensions{.pdf,.png,.jpg, .eps}

\begin{document}

  \title{Light Curve Properties of Supernovae Associated With Gamma-ray Bursts}
  
  \author{Xue Li 
  	\and Jens Hjorth}

	      \institute{Dark Cosmology Centre, Niels Bohr Institute, University of Copenhagen,   Juliane Maries Vej 30, 2100, Copenhagen, Denmark, \\
	      \email{lixue@dark-cosmology.dk}
        }

\date{Received date/Accepted date}

%===============================================================
% Abstract
%===============================================================

 \abstract
{Little is known about the diversity in the light curves of supernovae (SNe) associated with gamma-ray bursts (GRBs), including whether the light curve of SN 1998bw can be used as  a representative template or whether there is a luminosity-decline rate relation akin to that of SNe Ia.}
{In this paper, we aim to obtain well-constrained light curves of GRB-SNe without the assumption of empirical or parametric templates and to investigate whether the peak brightness correlates with other parameters such as the light curve shape or the time of peak.}
{We select eight   SNe in the redshift range 0.0085 to 0.606, which are firmly associated with GRBs.  The light curves of these GRB-SNe are well sampled across the peak.    Afterglow and host galaxy contributions are subtracted and dust reddening is corrected for.  
Low-order polynomial functions are fitted to the light curves.  A  K-correction is applied to transform the light curves into the rest frame V band. }
{ GRB-SNe   have   fairly uniform peak luminosities,  similar to SNe Ia.  Moreover, GRB-SNe follow a luminosity-decline rate relation similar to the Phillips relation for SNe Ia.  The relation between the peak magnitude $M_{\rm V,peak}$  and the   decline rate   $\Delta m_{\rm V,15}$ in V band  
 is  $M_{\rm V,peak} = 1.59^{+0.28}_{-0.24}  \Delta m_{\rm V,15} - 20.61^{+0.19}_{-0.22}$ mag,     
 with $\chi^2 = 5.2$ for 6 degrees of freedom.    
  This luminosity-decline rate relation is tighter than the $k-s$ relation,   
where $k$ and $s$ are the factors describing the relative brightness and width to the light curve of     SN 1998bw. 
The peak luminosities of GRB-SNe are also weakly correlated with the time of peak: the brighter the GRB-SN, the longer   the rise time.  }
{The light curve of SN 1998bw, stretched around the time of explosion,  can be used as a  template for  GRB-SNe with reasonable confidence, but stretching around the peak produces better results. GRB-SNe exhibit a luminosity-decline rate relation, similar to SNe Ia, both in normalization and slope. The existence of such a relation provides a new  constraint on GRB explosion models.  Considering the usefulness of SNe Ia in measuring cosmological distances, it is possible that GRB-SNe can be used as  standardizable candles to measure cosmological distances and constrain cosmological parameters.   }

               
     \authorrunning{Li \& Hjorth}
%     \titlerunning{Light Curve Properties of Supernovae Associated With  Gamma-ray Bursts}
   \titlerunning{Light Curve Properties of GRB-SNe}
 \maketitle

%===============================================================
%Introduction
%===============================================================
\section{Introduction}

Gamma-ray bursts (GRBs)  were first  observed by the Vela Satellites in 1967   \citep{klebesadel_1973}. 
Being flashes of narrow beams of intense electromagnetic radiation observed in distant galaxies \citep{metzger_1997}   with  peak energies  in the  gamma ray energy range, 
they are   the most luminous phenomena in the universe. 
The bursts are usually separated into two classes: long and short \citep{kouveliotou_1993}. The long GRBs have a duration of more than two seconds, while the short events last less than two seconds.     
Since the first discovery of the connection between SN 1998bw and GRB 980425  \citep{galama_1998bw, iwamoto_1998, kulkarni_1998, woosley_1998},     many   SNe    have  been found to be associated with long  GRBs \citep{hjorth_2003dh, stanek_2003dh, woosley_2006, hjorth_connection}.  %These SNe are characterized by relative smooth spectra and a very large explosion energy. 
 
%   The study on the association of the SNe and the GRB is developing very fast \citep{lixinli_connection, hjorth_2013}. 

   A collapsar model   \citep{MacFadyen_1999, MacFadyen_2001}  %\citep{fujimoto_collapsar}
      has been developed to   explain the GRB-SN connection.    But the properties of GRB-SNe, as well as GRBs are still under debate, e.g., what are the progenitors for long and short GRBs?       Can GRB-SNe be used as   standard candles?    To answer these  and other relevant questions, light curves of GRB-SNe are  required.

 In previous studies, the light curves   of SN 1998bw in the  U, B, V, R, I bands were used as a template to model the light curves of GRB-SNe  \citep{bloom_nature,  ferrero_2006aj, canoz_2013}.  The light curves of SN 1998bw were shifted   to the corresponding  redshift and scaled to the peak luminosity and stretched in time \citep{cano_2010bh}.  
Some work used   semi-analytical models %\citep{arnet_lc, jeffery_lc, richardson_2006} 
\citep{richardson_2006} to constrain GRB-SN light curves \citep{richardson_lc}.   But whether  the semi-analytical models or  the light curve of SN 1998bw can be used as a stretchable template is still unknown, not to mention if there is a better way to stretch the template other than stretching with   factor $s$  \citep{cano_2010bh}.   
To test this, it is important to obtain light curves instead of   using the SN 1998bw light curve as a template. 
 
It is not easy to obtain light curves of GRB-SNe. Sometimes a GRB is so bright that even though its afterglow declines rapidly  \citep{Paradijs_1997},  it may still exceed the brightness of its associated SN, in which case no SN will be detectable. This is also the case when a host galaxy is brighter than a GRB-SNe \citep{hjorth_2013}. Dust along the line of sight will extinguish the SN light.  Moreover,  any constraint  we impose on the light curves  e.g., afterglow modeling   or SN light curve modeling,  may bias the study. 
%  \citep{beuermann_afterglow, rhoads_afterglow}
  
 Our task is to find a way to obtain light curves of GRB-SNe without using light curve templates. With such light curves, we may  test if %GRB-SNe have constant luminosities or if their 
   peak luminosities of GRB-SNe are correlated  with other properties of the light curves, such as is the case for SNe Ia  \citep{phillips_1993, riess, phillip_1999}. We may further test if the light curve  of SN 1998bw can be used as a general light curve template for GRB-SNe, and,  if so,  how to stretch this light curve template.

The outline of this paper is as follows. In section \ref{sec:lc}, we discuss  the general steps in obtaining light curves of GRB-SNe from published data. In section \ref{sec:sys}, we   present the data and obtain   light curves  for  GRB-SNe.  Then  in   section \ref{sec:property}  we analyze the properties of the light curves of GRB-SNe. We     present  a     luminosity-decline rate  relation    and other properties of the light curves for GRB-SNe.  We also test if the light curve of SN 1998bw can be used as a general light curve template and if there is    a better  way  to stretch it than the commonly used approach. 
In section \ref{sec:summary},  we summarize our investigation and discuss future prospects.

 %%%%%%%%%%%%%%%%%%%%%%%%%%%%%%%%%%%%%%%%%%%%%%%%%%%%%%%%%%%%%%%%%%%%%%%

\section{Light curves of GRB-SNe}
\label{sec:lc}

  Our goal is to obtain   light curves of GRB-SNe in the rest frame V band. This is because, as shown in Figure~\ref {template}, the spectral energy distribution (SED) peaks around the V band. In addition, the K-correction procedure (section~\ref{subsec:totalk})  relies on using a redder band to correct to the rest-frame flux and we rely on the availability of suitable data. 
  After the light curves are obtained, we      measure   peak magnitudes  ($M_{\rm V,peak}$)    and   decline rates ($\Delta m_{\rm V,\alpha}$). 
   Here $\Delta m_{\rm V,\alpha}$  is defined as the decline of the V-band magnitude $\alpha$ days   after the SN has reached its peak brightness.  
   
   For error estimation, we use a standard Monte Carlo method to resimulate       the data throughout the paper. 
The resulting uncertainties are obtained as 68.3\% $ (\pm 1  \sigma) $ of the total resimulated results.  
 In general, to obtain a light curve, we  account for the effects of the host galaxy and the afterglow,  and subtract their contributions from the total flux.        We correct for extinction and fit  low-order polynomial functions to the resimulated   data. %We calculate  distance modulus to convert   apparent magnitude into   absolute magnitude.  
   A K-correction  is used to get the peak magnitude and decline rate in the rest frame V band. To do so, we either   apply a multi-band K-correction, or  use the
 SN 1998bw peak SED and decay properties to correct the values of the peak magnitude and the decline rate in bands obtained in a wavelength close to the rest frame V band.

\subsection{Host galaxy}
\label{subsec:hostgal}

The brightness observed is the total flux of the GRB-SN, the afterglow,   and    the host galaxy. In some cases, the host galaxy is sufficiently faint compared to the SN that the host   contribution is negligible. But for other systems, the host galaxy will contaminate the SN light curve. In these cases, to obtain the intrinsic SN luminosity, the contribution of the host galaxy must be subtracted.

The brightness of a host galaxy is constant.  It is usually observed   when the SN  has faded away. %Sometimes the host flux is estimated by fitting it jointly with a template SN light curve.  
In this paper, we take the host brightness from   the  literature.    
  The host brightness is resimulated with the standard Monte Carlo method,  and subtracted from  the  total brightness.

\subsection{Afterglow}
\label{subsec:afglow}

\begin{table*}[h!tbp]
\begin{center}
\caption{The slopes and the break times for GRB 050525A and GRB 090618.  }
 \label{tab:slope}
% \begin{tabular}{|c|cccc|cc|c|}
 \begin{tabular}{c|cccc|cc|c}
  
    \hline
 \hline
   GRB/XRF/SN &   \multicolumn{4}{ c|}{   smooth function$^a$}  &   \multicolumn{2}{ c |}{   broken power-law} &  reference    \\ %\tnote{1} \\ 
%            &             &               & subtraction              &    subtraction              &                        &    &    \\
                          &                          $\bar t_{\rm break}$ (day) & $\bar  \beta_1$ & $\bar \beta_2$ & $\widehat \beta$$^b$&  $t_{\rm break}$ (day) &  $  \beta_2$  &   \\
  % slope &  $t_{\rm break$ 
  \hline

  % 050525A/2005nc   &     (1), (2)  & $0.3 $ & 1.1  & 1.8  & 1.57 & 0.3   &   $ 1.57 ^{0.07}_{0.08}$ \\
 %  050525A/2005nc    & $0.3 $ & 1.1  & 1.8  & 1.57 & 0.3   &   $1.74^{0.11}_{0.12}$   &     (1), (2)\\
    050525A/2005nc    & $0.3 $ & 1.1  & 1.8  & 1.63 & 0.3   &   $1.74^{+0.11}_{-0.15}$   &     (1), (2)\\
 % 090618    &     $0.48 \pm 0.08$ & $0.79 \pm 0.01$ & $ 1.74 \pm 0.04$ & $1.43^{0.06}_{0.06}$  &  0.5 & $ 1.37 ^{0.11}_{0.13}$ &  (3)   \\
  090618    &     $0.48 \pm 0.08$ & $0.79 \pm 0.01$ & $ 1.74 \pm 0.04$ & $1.52^{+0.06}_{-0.05}$  &  0.5 & $ 1.56 ^{+0.07}_{-0.08}$ &  (3)   \\
  %$ 1.57 ^{0.07}_{0.08}$
  \hline
% \hline
                                                                   
\end{tabular}
\end{center}
{\footnotesize
 \noindent
 
 \noindent
$^a$:  An afterglow fitting method with        $m (t) = -2.5 \times \left( \left( \left( \frac{t}{t_{\rm break}}  \right)^{\bar \beta_1} + \left(\frac {t}{t_{\rm break}}  \right)^{\bar \beta_2} \right)^{-1} \right) + B $     \citep{cano_060729}.  \\

\noindent
$^{b}$: Post-break slope fitted to the smooth function with broken power-law method.  \\

 (1) \cite{blustin_050525},    (2)  \cite{dellavalle_050525},  
(3) \cite{cano_060729}}

\end{table*}

 Except for two long GRBs, i.e., GRB 060614 \citep{fynbo_nature, galyam_060614, gehrels_060614} and GRB 060505 \citep{fynbo_nature,ofek_2006, mcbreen_060505}, for which no   associated SNe were detected,  there are no other known cases of long duration GRBs for which the limits on detecting a SN rules out something that is about as bright as SN 1998bw \citep{hjorth_connection}.

   GRBs  are very luminous and energetic with isotropic energies   up to  $E_{{\gamma, iso}}  \sim  10^{54}$ erg  \citep{hjorth_connection, xu_2013}.  Soon after the burst, the flux of  the  GRB afterglow dominates the light, but it declines rapidly.   In some cases, after a few days,  the brightness of an afterglow will have decreased significantly and is no more a major  contributor to the photometry. At this time, if the host galaxy is not brighter than the SN, usually we can observe the light from the SN.

We assume that the afterglow behaves as a power law or a broken power-law decay      $f (t) = c_1  t ^{\beta_1}$ for $t < t_{\rm break}$ and  $f (t) =  c_2    t ^{\beta_2}$ for $t \geqslant t_{\rm break}$, where $f(t)$ is the flux of an afterglow, $\beta_1$ and $\beta_2$ are the decay slopes and  $ t_{\rm break}$ is the time for the  change of the slopes   from pre-break $\beta_1$ to post-break $\beta_2$.    We choose $ t_{\rm break}$ based on the data or   from the literature.  This method is different from the     broken power-law fits  \citep{zeh_afterglow, cano_060729}, but  the effect is similar. The slopes, the break time for systems GRB 050525A and GRB 090618 (see section \ref{sec:sys} for more discussion on each system) are listed in Table \ref{tab:slope}. The values  of  $\bar t_{\rm break}$,   $\bar  \beta_1$ and $\bar  \beta_2$ in the 
column of  `smooth function' are   from the literature, while in the  `broken power-law' column,   the value of $\beta_2$ is used in this paper.  When the results of the smooth function are fitted with a broken  power-law way,  $\widehat \beta$ is the  fitted post-break  slope.  Compare $\widehat \beta$ and $   \beta_2$, we conclude that the broken  power-law fits is consistent with the results of broken power-law fits.

   In this paper,  we resimulate the afterglow data with the standard Monte Carlo method and fit the resimulated data with broken power-law functions. %if not mentioned, we resimulate    the afterglow decay with the standard Monte Carlo method. Otherwise, we fix the   decay slope    from values  given in the literature  obtained from      X-ray observations of a afterglow.      
We remove the   contribution of  the afterglow by subtracting the fitted power-law functions.

   \subsection {Extinction, distance modulus,  and  rest frame time}
 \label{subsec:otheffect}

Dust in galaxies reddens   light emitted from  GRB-SNe.   There are two  main contributing sources: dust in the   host galaxy,   where the GRB-SN  is located,   and dust in  the  Milky Way.  In this paper,  we take the values of host extinction, e.g.,    $A(V)_{\rm host}$ or  $E(B-V)_{\rm host}$,    from   the literature.    
        \cite{schlafly_dustmap}    found      that      the Galactic extinction  is overestimated by  DIRBE/IRAS   dust map \citep{schlegel_dustmap} and calculated the correction coefficients if the dust map is used.  % and calculated the   re-calibration  factors in bands. 
In this paper,  with  $R_V = 3.1$,   we use the  DIRBE/IRAS   dust map \citep{schlegel_dustmap} to get $E(B-V)$, then the coefficients (see Table 6 in    \cite{ schlafly_dustmap}) are multiplied to correct the value. 
 
   % IRAS/ISSA{schlegel_dustmap,schlafly_dustmap}  to estimate its effect.  A value of  $R_V = 3.1$ is applied and the value  of  $A (\lambda)$ is   based on  the ratio  of $A (\lambda) / A(V)$  \citep{schlafly_dustmap}. 
 % to estimate its effect.  
% and the value  of  $A (\lambda)$ is   based on  the ratio  of $A (\lambda) / A(V)$  \citep{schlafly_dustmap}. 
    
 % For the foreground extinction, we apply  IRAS/ISSA  dust map \cite{schlegel_dustmap} to estimate its effect.  The value of  $R_V = 3.1$ is applied and the value  of  $A (\lambda)$ is   based on  the ratio  of $A (\lambda) / A(V)$ \cite{cardelli_1998}. 

 The distance modulus is calculated and subtracted to obtain the absolute magnitude.   In this paper, we adopt the    cosmological parameters $\{ \Omega_m, \Omega_{\Lambda}\} = \{0.315, 0.685\}$ and  $H_0 = 67.3$   km s$^{-1}$  Mpc$^{-1}$   in a flat universe \citep{planck_cosmopara}.  The absolute magnitude    is determined as
  \begin{equation}  
%    m_{lensed} = M + 5 \log_{10}( D_{L} /10 {\rm  pc}) - 2.5 \log_{10}(|\mu|) -2.5 \log_{10}(1+z).
     M = m -  5 \log_{10}( D_L /10 {\rm  pc}) - \Delta K  + 2.5 \log_{10}(1+z), 
      \label{magthres}
\end{equation} 
 \noindent where $D_L$ denotes the luminosity distance and  $\Delta K$ represents the effect of the K-correction (section \ref{subsec:totalk}).

Peculiar velocities may affect the estimate of the distance modulus, especially for nearby SNe.  Except for SN 1998bw which   has peculiar velocity  $ v_p = - 90 \pm 70 $ km s$^{-1}$ \citep{li_cos},  for the other systems we assume the peculiar velocity is $0$ and the uncertainty is   $\delta v_p = 300$ km s$^{-1}$    \citep{tamara_pec}.    
Therefore, the uncertainty in the distance modulus is   $(5/2.3) \,  \delta v_p \, (c z)^{-1}$,     where $c$ is the speed of light and $z$ is the redshift of a GRB-SN.

 It is straightforward to convert  the observational time into  the  rest frame time  by  dividing the observational time by $(1+z)$.

 \subsection{Polynomial function fitting}
 \label{polyfunction}
 
After the steps discussed above,  polynomial functions are fitted to the data. We fit the data with the lowest possible order. The most suitable order is to 
some extent subjective, but as discussed below, in most cases the polynomial functions are of 3rd or 4th order. Data on both sides of the peak are needed.  %This is to make sure the fitted polynomial functions represent real SN light curves.
This is to ensure that parameters dependent on sampling the peak, such as the time of peak, the peak brightness, and the decline rate past peak, are robustly determined.

 \subsection{K-correction}
 \label{subsec:totalk}

The observational data of   GRB-SNe     may be in U, B, V, R, I and other bands. After subtracting the host and afterglow brightness  and fitting  polynomial functions to the results,       light curves of   SNe are obtained in the observed    bands.  Then a K-correction is applied to correct   light curves from   the observed band(s) into the rest frame V band. 
If the   systems have been observed in two  or more  bands and   two bands are close to the redshifted V band  for  interpolation, then a `multi-band K-correction'  \citep{vandokkum_kcorrection, hogg_kcorrection}  %\citep{hogg_kcorrection, vandokkum_kcorrection, fukugita_1995}
 is applied.  For the other systems,  which have data in only one band   or the other bands are not close to the redshifted V band, we correct their peak magnitude and decline rate with    peak SED and decline rate templates based on SN 1998bw.  We use broad-band data because we do not have useful 
spectra around (before and after) the peak of the light curve.

  \subsubsection {Multi-band K-correction}
 \label{subsec:kcorrection}

   The multi-band K-correction  is a method to constrain the  light curves in the rest frame V band by interpolating two light curves in adjacent observed bands   \citep{vandokkum_kcorrection, hogg_kcorrection}.  
    The method is based on the assumption  that    flux densities are correlated in contiguous bands. For example, if a GRB-SNe has $z \in (0.26, 0.60)$, then we can   interpolate the magnitude in the R and I bands into  the redshifted V band. The %flux density in the V band can be estimated  as: $F(\nu_{V}(z)) = F(\nu_{R})^{\alpha} F(\nu_{I})^{1-\alpha}$.
flux density in the V band can be estimated  as: $F(\nu_{V}(z)) = F(\nu_{R})^c F(\nu_{I})^{1-c}$.
 The magnitude in the redshifted V band is  then
 %%%%%%%%%%%%%%%%%%%%%%%%%%%%%%%%%%%%%%%%%
     \begin{equation} 
V_z =   c  R  + (1 -c)  I ,
  \label{kfunction}
\end{equation}  
 %%%%%%%%%%%%%%%%%%%%%%%%%%%%%%%%%%%%%%%%%
  \noindent where $V_z$,   $R$ and $I$ are magnitudes in the AB system.   The parameter $c$ is calculated as a function of   central wavelength of the observed bands and the SN redshift \ \citep{vandokkum_kcorrection}. Here $c =(\lambda_I -    \lambda_V (1+z) )/(\lambda_I - \lambda_R)$  with $\lambda_R$  and $\lambda_I$ being the observational R and I band wavelengths and      $   \lambda_V (1+z) $ being the redshifted V band wavelength.     In this step, the selected two bands should   fulfill  the conditions: 1) the two bands must be adjacent; 2) the parameter should be $c \in (0,1)$ to make sure one is not extrapolating beyond the observed bands.

\subsubsection{SN 1998bw peak SED and decline rate templates}
\label{subsec:template}

\begin{figure}

\begin{center}

\includegraphics[width=8.8cm] {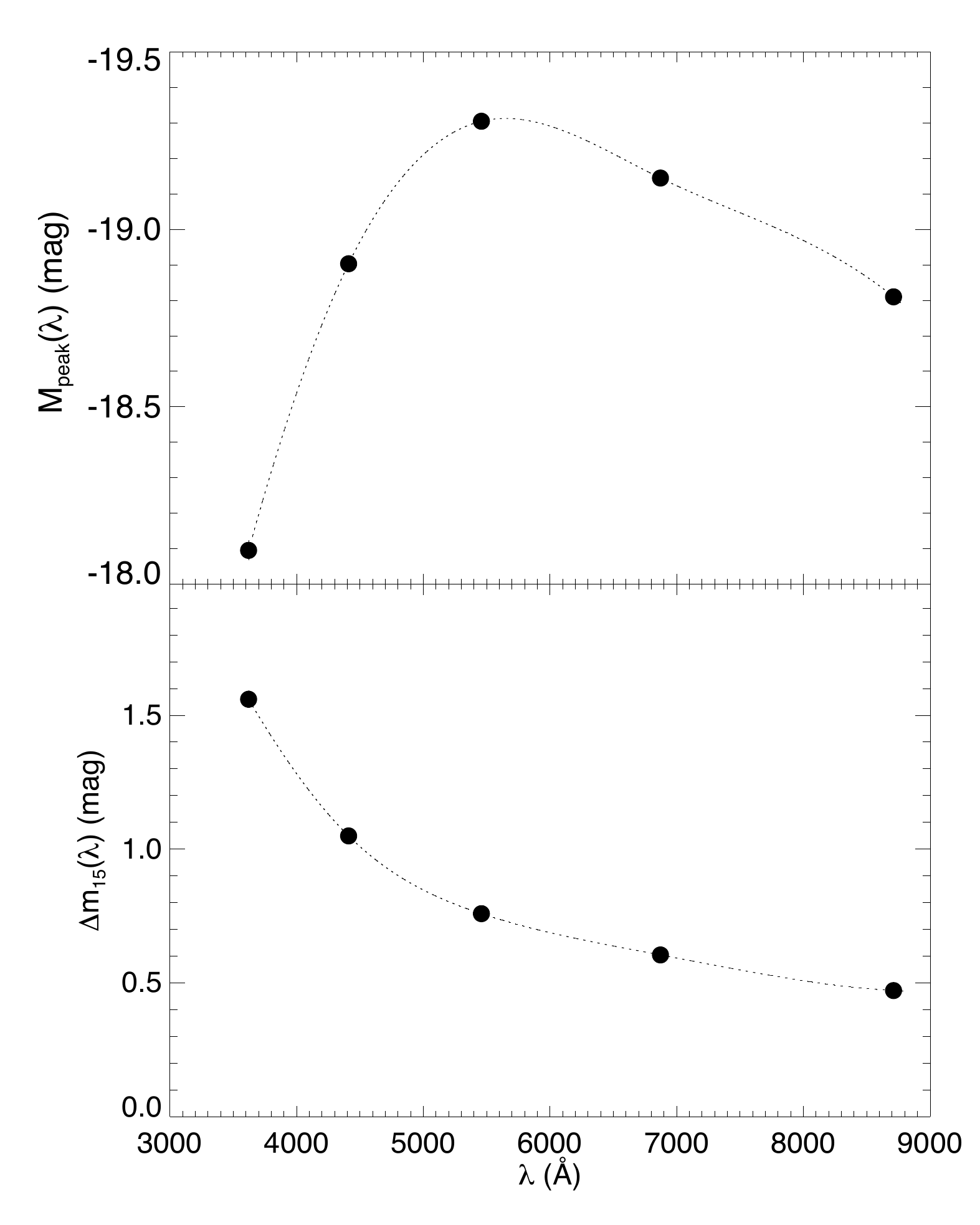}

\caption{   SN 1998bw peak SED and decline rate templates. The black points represent  values in the U, B, V, R, I bands   \citep{galama_1998bw, clocchiatti_1998bw, sollerman_1998bw}. The dotted lines are spline interpolations. The upper panel shows the relation between the wavelength  $\lambda$  and the peak magnitude   $M_{\rm peak}(\lambda)$. In the lower panel, the relation between the wavelength  $\lambda$  and the decline rate  $\Delta m_{ \alpha}(\lambda)$  is   plotted. Here we show values for $\alpha = 15$.  
 $M_{\rm peak}$ values have errors $<0.02$ mag while $\Delta m_{15}$ values have errors $< 0.026$ mag in    the   V, R and I bands. 
 }
  \label{template}
\end{center}
\end{figure}

   \begin{figure}
\begin{center}
\ 
\includegraphics[width=8.8cm] {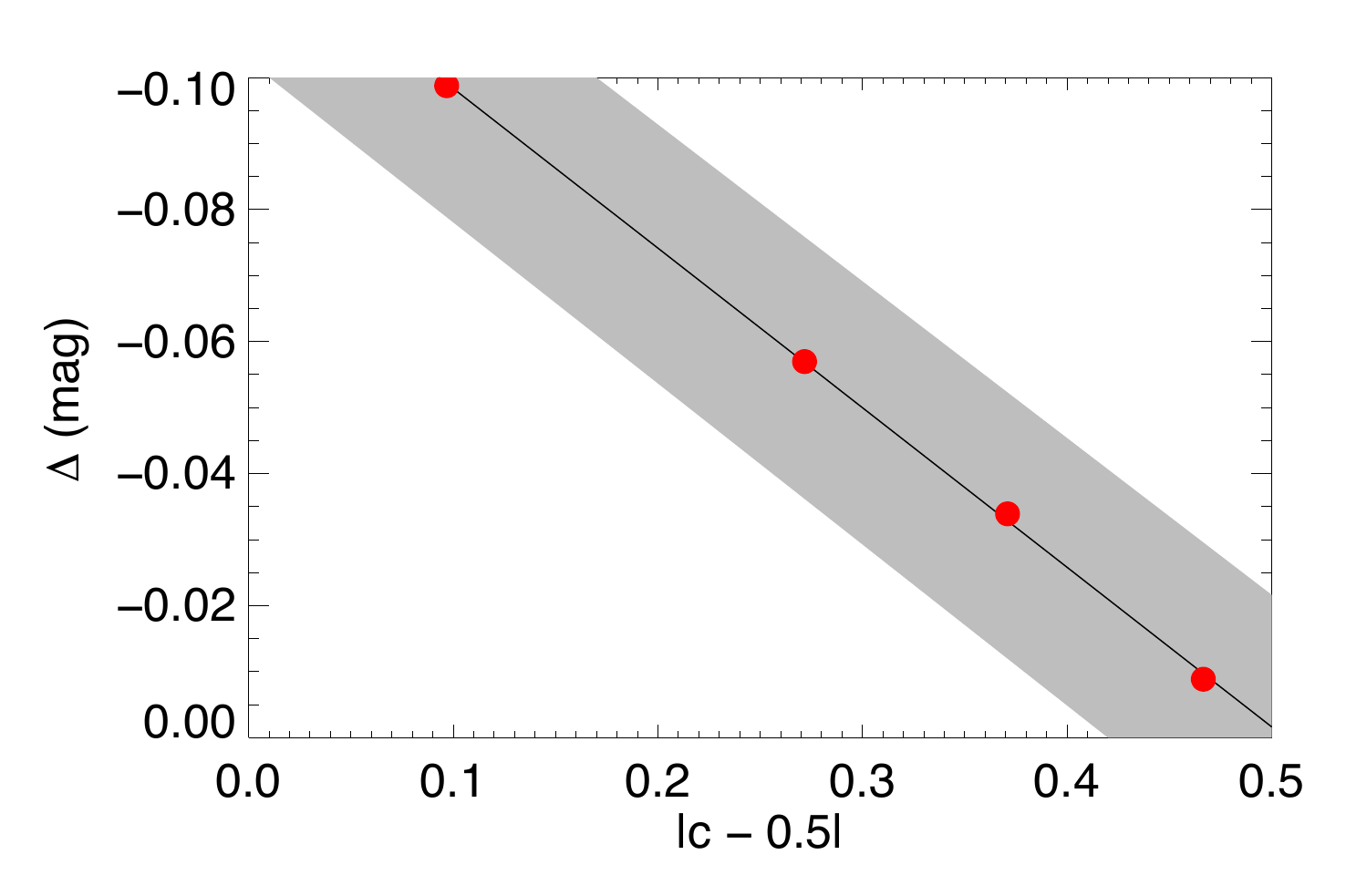}

\caption{ Difference of peak magnitude  ($\Delta$) estimated via  two K corrections %based on the multi-band K-correction and  the SN 1998bw templates 
as a function of K-correction factor $c$  (Eqs. \ref{kfunction} and \ref{deltac}). We show the results for four systems: SN 1998bw, SN 2006aj, SN 2010bh and SN 2012bz  (see section  \ref{sec:sys} for details).  The shaded area shows the systematic uncertainty of 0.02 mag  in the K  correction procedure. }
  \label{syserr}
\end{center}
\end{figure}

%%%%%%%%%%%%%%%%%%%%%%%%%%%%%%%%%%%%%%%%%%%%%%%%

When useful light curve data is available only in one band, or the other observed  bands are too far away from the redshifted V band to do a meaningful 
multi-band K-correction,  we resort to using   the light curves of  
SN 1998bw as a template to obtain
V-band values from data close to the redshifted V band, typically within a few hundred \AA. 
The observed    band which is closest  to the  redshifted V band is chosen  to obtain the light curves and measure the values of the peak magnitude    and   the decline rate. After that,    according to the wavelength of the chosen band, $M_{\rm V,peak}$  and $\Delta m_{\rm V,\alpha}$ are corrected to the rest frame V band using   the light curves of    SN 1998bw as a template.   

The SN 1998bw peak SED and decline rate templates      describe   the relations  of       $M_{\rm peak}$ and    $\Delta m_{alpha}$, as a function of   wavelength $\lambda$.  It  is  based on  the assumption that  %\sout{  light curves in a certain band behave similarly  for all GRB-SNe.}  
  the behavior of the  light curves in different bands are similar for all GRB-SNe.   
  Compared to other effects, e.g., extinction, host and afterglow  subtraction, this is a 2nd order effect and does not require that the overall light curves or spectra are perfectly 
identical to those of SN 1998bw.

Here we use the observational   data of   SN 1998bw  \citep{galama_1998bw, clocchiatti_1998bw, sollerman_1998bw}  to establish the    template (see section \ref{subsec:1998bw} for details on SN 1998bw). 
 The light curves   are well defined by the observational data.  Therefore,     peak magnitudes   $M_{\rm peak}$  and the decline rates  $\Delta m_{\alpha}$    are constrained in the U, B, V, R, I bands.   
Then   we spline interpolate the relations between  $M_{\rm peak}$,  $\Delta m_{\alpha}$ and $\lambda$.  The resulting
templates  are shown in Figure  \ref{template}.    %For the ones whose peak magnitude $M_{\lambda}^{\rm data}$ and decline rate $\Delta m_{\rm V,\alpha}^{\rm data}$  need to be corrected with these templates, on the template 
%\sout{ We correct the peak magnitude by  measuring   $M_{V}^{\rm temp}$ and $M_{\lambda}^{\rm temp}$  at the wavelength $\lambda = \lambda_{\rm obs}  /(1+z)$, with $\lambda_{\rm obs}$ being the observational   wavelength.    
%The peak magnitude is corrected by     $  M_{\rm V, peak} = M_{\lambda}^{\rm data}  - M_{\lambda}^{\rm temp}    +  M_{V}^{\rm temp}$.  }
The peak magnitude is corrected as     $  M_{\rm V, peak} = M_{\lambda}^{\rm data}  - M_{\lambda}^{\rm temp}    +  M_{V}^{\rm temp}$, where $M_{\lambda}^{\rm data} $ represents the peak magnitude at the wavelength $\lambda = \lambda_{\rm obs}  /(1+z)$, with $\lambda_{\rm obs}$ being the observational   wavelength, while  $M_{\lambda}^{\rm temp}$  and  $M_{V}^{\rm temp}$  denote the measured values at the wavelength   $\lambda$ and in the V band respectively, obtained from the SN 1998bw peak SED and decline rate templates. 
 The decline rate $\Delta m_{\rm V,\alpha}$   is corrected in the same way.   % where i represents the value at the rest frame V band wavelength $\lambda_v$ and j is the values at the wavelength $\lambda_{\rm obs} /(1+z)$, with $\lambda_{\rm obs}$ being the observed band wavelength.   

 The differences of $M_{\rm V,peak}$ estimated via the two methods of K-correction are shown in  Figure  \ref{syserr} for four systems: SN 1998bw, SN 2006aj, SN 2010bh and SN 2012bz (see section \ref{sec:sys} for more discussion on each system). Here  $|c - 0.5|$ represents the relative  distance of the redshifted V band wavelength to the average of two observed wavelengths. The differences between the two methods may get bigger as the redshifted V band is closer to the middle point of two observed bands.  
Here    $\Delta = M_{\rm V,peak}^{\rm temp} - M_{\rm V,peak}^{\rm multi}$, where $M_{\rm V,peak}^{\rm temp}$  denotes the peak magnitude estimated via the  SN 1998bw peak SED and decline rate template    and $M_{\rm V,peak}^{\rm multi}$ is the value obtained from the multi-band K-correction.     A linear fit to  the relation gives
      \begin{equation} 
 \Delta  =  0.24 \cdot  |c - 0.5| - 0.12. 
   \label{deltac}
\end{equation}  
 %%%%%%%%%%%%%%%%%%%%%%%%%%%%%%%%%%%%%%%%%

  Figure \ref{syserr}  shows that two methods lead to small differences with $|\Delta|  \leqslant$ 0.1 mag.  The peak of $M_{\rm peak}(\lambda)$ on the SN 1998bw peak SED and decline rate templates   is around 5500 $\AA$, which is close to the V band. Therefore, the multi-band K-correction, obtained by interpolating magnitudes on each side of the peak, may underestimate   the value of $M_{\rm V,peak}$.   We therefore apply the modification
  (Eq. \ref{deltac}) to the multi-band K-correction. We adopt a      systematic uncertainty of   0.02    mag in quadrature in the K correction after applying this
  modification, as shown in the shaded area in Figure \ref{syserr}.

\section{Systems of GRB-SNe}
\label{sec:sys}

\begin{table*}%[h!tbp]
\begin{center}
\caption{The selected systems and the relevant steps. }
 \label{listgrbsn}
 \begin{tabular}{ccccccc} 
    \hline
 \hline
   GRB/XRF/SN &      afterglow$^{a}$     & host$^{b}$     & k/t$^{c}$ & class     & reference          \\ %\tnote{1} \\ 
%            &             &               & subtraction              &    subtraction              &                        &    &    \\
  \hline

980425/1998bw       & - & - &$k$ & $A$ &    (1), (2), (3)  \\
%   && &&\\
% \hline
030329/2003dh   & -& -& $t$& $A$ &  (4), (5) \\
% \hline
 031203/2003lw    & - &$\surd$ &$t$ & $A$ &   (6), (7), (8)   \\
% \hline
  050525A/2005nc    & $\surd$ &  $\surd$ & $t$ & $B$ &     (9), (10)   \\
% \hline
 060218/2006aj     &- & $\surd$ & $k$ & $A$ &  (11),  (12), (13), (14), (15) \\
% \hline
  090618   & $\surd$ &  $\surd$ & $t$ & $C$ &       (16) \\
 %  \hline
 100316D/2010bh   & - & - &$k$  & $A$&  (17),  (18), (19)  \\
%  \hline
 120422A/2012bz     &$\surd$  & - &$k$ & $A$ &   (20),  (21)  \\
  \hline
% \hline
                                                                   
\end{tabular}
\end{center}
{\footnotesize
 \noindent
 
 \noindent
$^a$: Subtraction of   afterglow brightness.  

\noindent
$^{b}$: Subtraction of   host galaxy    brightness. 

\noindent
$^{c}$: Multi-band K-correction (denoted \textquoteleft k') or shift based on the SN 1998bw peak SED and decline rate templates (denoted \textquoteleft t'). \\
(1) \cite {galama_1998bw}, (2) \cite{sollerman_1998bw},  (3)  \cite{clocchiatti_1998bw},  
(4) \cite{hjorth_2003dh}, (5) \cite{matheson_2003dh}, 
 (6) \cite{malesani_2003lw_one},  (7) \cite{mazzali_2003lw},  (8) \cite{malesani_talk}, 
(9) \cite{blustin_050525},    (10)  \cite{dellavalle_050525},  
(11) \cite{sollerman_2006aj},  (12) \cite{ferrero_2006aj}, (13) \cite{simon_2006aj},    (14) \cite{guenther_2006},  (15) \cite{poznanski_extinction}, 
(16) \cite{cano_060729},  
(17)  \cite{cano_2010bh},   (18) \cite{olivares_e_2010bh},   (19)  \cite{bufano_2010bh}, 
(20)  \cite{melandri_2012bz},  (21) \cite{schulze_2012bz}.   
}

\end{table*}

Based on the degree of observational evidence of a GRB having an associated SN  \citep{hjorth_connection},   GRB-SNe are graded  from class  {\it A} to class {\it E},  where class {\it A}  has   `strongest spectroscopic evidence',   while class {\it E} has the weakest evidence. In this paper, we select  GRB-SN  candidates in classes {\it A}, {\it B}, and {\it C}.   In total 19 systems are collected  (see Table 9.1 in \citealt{hjorth_connection}), including SN 2012bz \citep {schulze_2012bz} which is classified as a class {\it A} system.

We   have studied all the 19 GRB-SN  systems  and evaluated the feasibility of   constraining model-independent peak magnitudes and decline rates for them.  Among them,  we succeed to measure $M_{\rm V,peak}$ and $\Delta m_{\rm V,\alpha}$  for    8 systems,   as 
 listed in  Table \ref{listgrbsn},  along with the corrections made in each case.  % X-ray Flash \citep{heise_2001} is denoted by XRF. 
 We address the other systems and the reasons why they are not selected in section \ref {noincludegrb}.

 \subsection{GRB 980425/SN 1998bw}
\label{subsec:1998bw}

%%%%%%%%%%%%%%%%%%%%%%%%%%%%%%%%%%%%%%%%%%%%%%%
   \begin{figure*}
 \centering
\includegraphics [width=17.6cm]{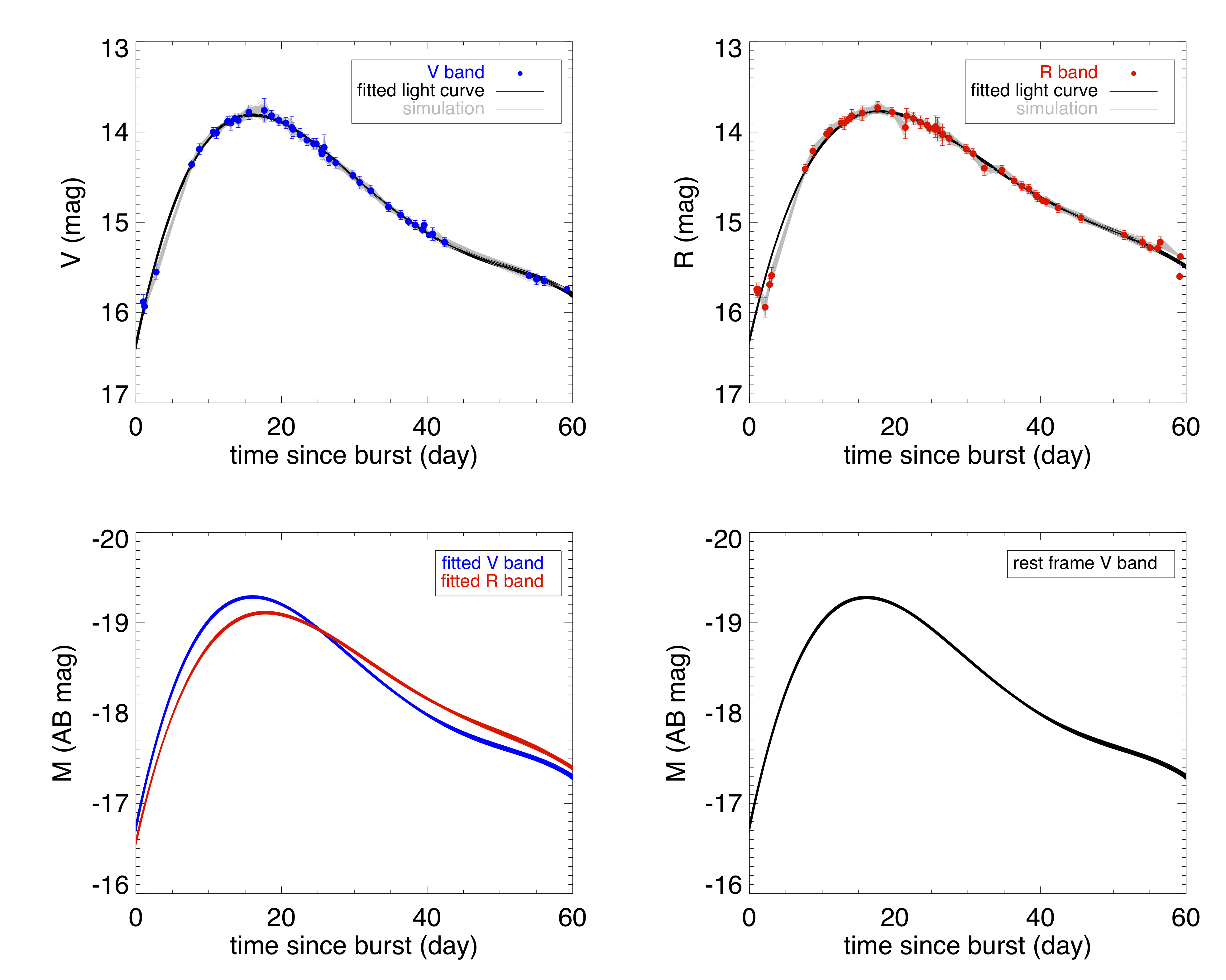}%{sn1998.pdf}
\caption{ GRB 980425/SN 1998bw.   The upper   panels  show   the photometric data points (blue/red), the resimulated   data (gray) and the   polynomial functions fitted to the resimulated   data in V ({\it   left}) and R   ({\it   right})  bands.   In the lower left panel, the light curves  after the extinction and the distance modulus correction are plotted in V band (blue) and R band (red) in AB magnitude. In the lower right panel, the final light curves after the K-correction in the rest frame V band are plotted.    
The   uncertainties in  the resimulated data and the fitting light curves are  plotted  as 68.3\% $ (\pm 1  \sigma) $ of the total resimulated results.   
 %\emph {\textbf{Throughout:}} for the similar figures in this section, the upper panel(s) is apparent magnitude whereas the lower panel(s) is absolute magnitude (in AB system). 
%\emph {\textbf{Throughout:}} the x-axis is rest-frame time. Photometric data and the host brightness are in  blue/red. resimulated   data   are in   grey. Polynomial functions fitted to the SN light curves are in black.
}
  \label{sn1998}
\end{figure*}
%%%%%%%%%%%%%%%%%%%%%%%%%%%%%%%%%%%%%%%%%%%%%%%%

SN 1998bw was the first   SN discovered to be  connected with a GRB, GRB 980425  \citep{galama_1998bw}.  % of any GRB-SN,
%$z = 0.0085$. 
%\textcolor{red}{
 % $z = 0.009$   \citep{foley_98bw}.      citation peculiar velocity citation from email }
Combined with the peculiar velocity   $v_p = - 90 \pm 70 $ km s$^{-1}$    \citep{li_cos}  and the CMB velocity $v_{\rm CMB} = 2505  \pm   14$  km s$^{-1}$  \citep{fixsen_cmb},  it has  $v_z = 2595 \pm 71$  km s$^{-1}$. 
The redshift is  
  $z =  v_z / c = 0.00866 \pm 0.00024$.   %\citep{fixsen_cmb, mould_2000, foley_98bw}.  
   It is by far the lowest redshift of GRB-SNe.   
 A lot of work have been done on this system, which is why we
have built the peak SED and decline rate templates based on this system.  
This system is a class $A$ GRB-SNe.  
 
We collect the data in the V and R bands    \citep{galama_1998bw, sollerman_1998bw, clocchiatti_1998bw}.  
    %There is no discussion in the literature on  the contribution of the brightness of  the host galaxy or the afterglow.  So 
    We assume that the host galaxy and afterglow contributions to the total brightness are negligible and fit 4th order polynomials to the light curves.
 We have enough data in the V and R bands and these two bands are close to the redshifted V band, so the multi-band K-correction is applied. The parameter    in the K-correction is $c=  0.97$. 
 We assume that the
 host extinction is negligible.      Unless stated otherwise, we treat GRB-SNe in the same way and neglect the contributions from the host galaxy, the afterglow or the host extinction,  
 if they are not mentioned in the literature.      %Polynomial functions with low degrees are fitted   to the resimulated   data. 
    The Galactic extinction is estimated to be  $E(B-V) = 0.06$ mag.  The distance modulus is     $\mu = 32.94 \pm  0.08$.    
     The uncertainty in the distance modulus   is dominated by the uncertainty in the  peculiar velocity.

Figure~\ref{sn1998}  shows the   light curves of SN 1998bw.  In the upper panel, the observational data,  the resimulated   data   and the fitting functions to the resimulated   data in the V   (left) and R  (right) bands are plotted.   In the lower panel, the left plot shows the light curves in the V and R bands, after correcting for the Galactic extinction and converting into the absolute magnitude. On the  right, after the K-correction,  the   light curves in the rest frame V band are  plotted.  The temporal axes in the four panels have been corrected into the rest frame.   
 The   uncertainties in  the resimulated data and the fitting light curves are  plotted  as 68.3\% $ (\pm 1  \sigma) $ of the total resimulated results.   
   In this section, the similar figures show 100 out of the 1000 resimulated        light curves.      %In   this section, we treat other systems in the same way, if not mentioned.

%%%%%%%%%%%%%%%%%%%%%%%%%%%%%%%%%%%%%%%%%%%%%%%%]

%%%%%%%%%%%%%%%%%%%%%%%%%%%%%%%%%%%%%%%%%%%%%%%%]

\subsection{GRB 030329/SN 2003dh}
 
  %%%%%%%%%%%%%%%%%%%%%%%%%%%%%%%%%%%%%%%%%%%%%%%%

\begin{figure*}
 \centering
\includegraphics[width=8.8cm] {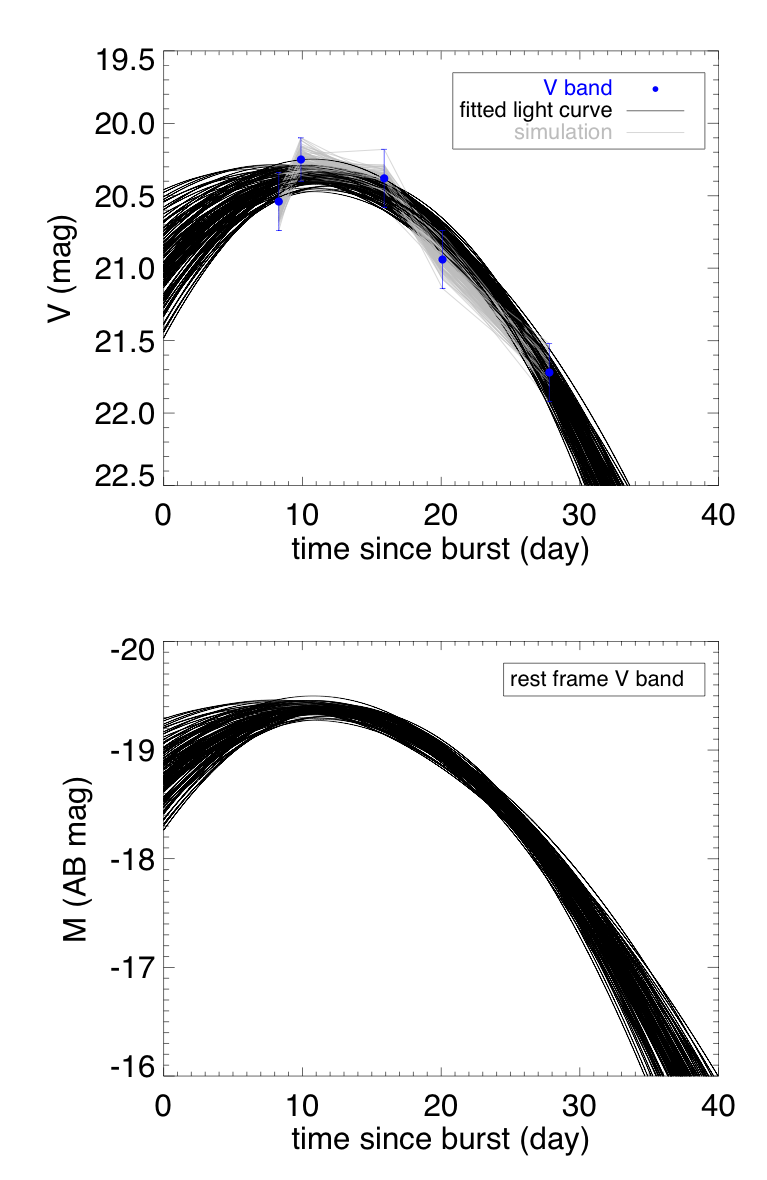}
\includegraphics[width=8.8cm]{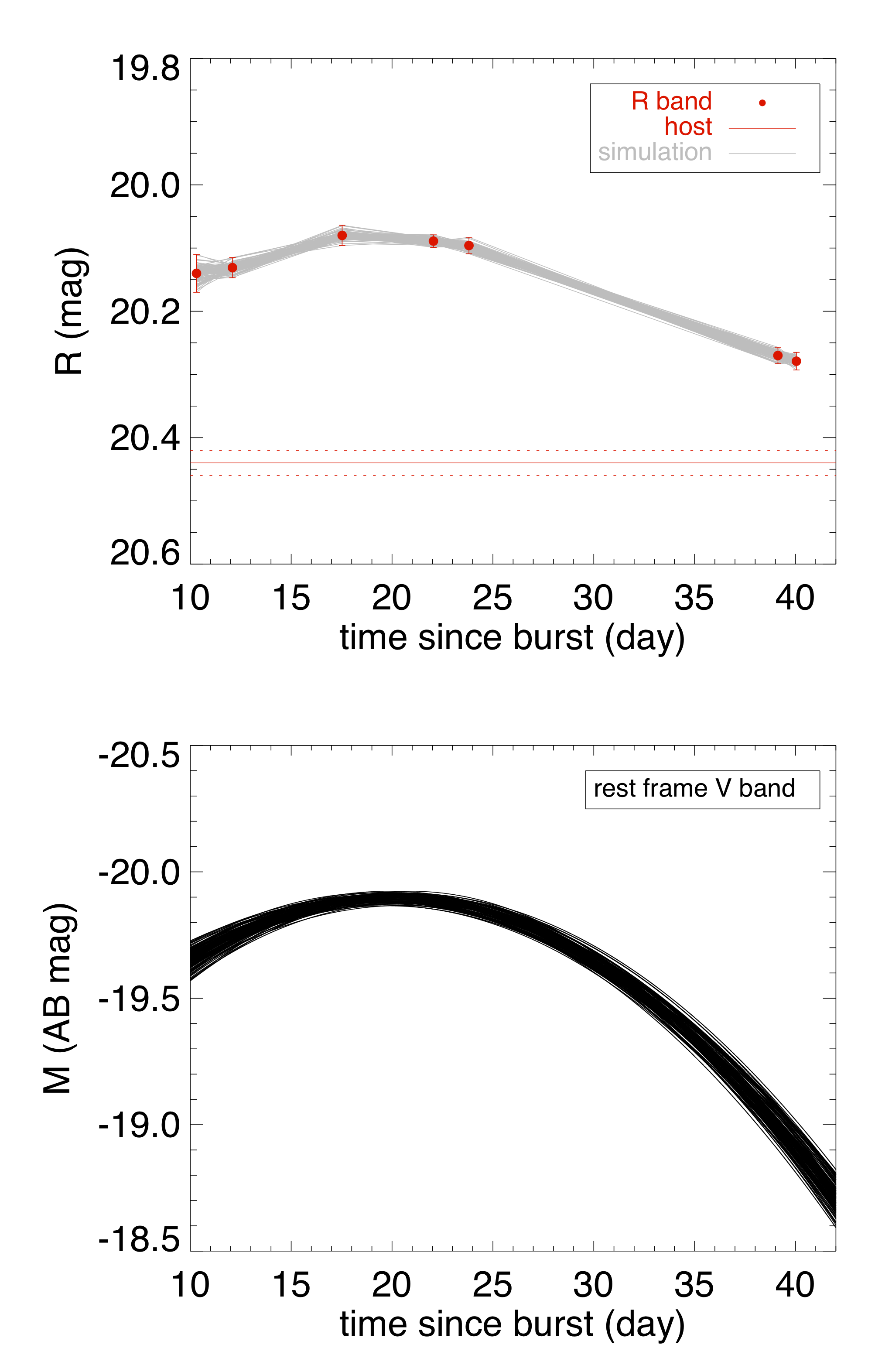}
\caption{GRB 030329/SN 2003dh ({\it   left}) and GRB 031203/SN 2003lw  ({\it   right}).  The line styles are    the same  as in Figure \ref{sn1998}.    
 %The   uncertainties in  the resimulated data and the fitting light curves are  plotted  as 68.3\% $ (\pm 1  \sigma) $ of the total resimulated results.   
  }   
\label{fig:sn2003dh}

\end{figure*}

The GRB 030329/SN 2003dh system was the first solid spectroscopic association
between a cosmological GRB and a SN \citep{hjorth_2003dh, stanek_2003dh}. 
The redshift is
$z = 0.1685$. It is a class $A$ system. The data is collected in the V band  \citep{hjorth_2003dh}.  With an SMC extinction law, \cite{matheson_2003dh}  estimated  the host extinction to be $A_{\rm V, host} = 0.12 \pm 0.22$ mag. We assume no host extinction,  which is consistent with the result from \cite{matheson_2003dh}. 
The Galactic extinction is $E(B-V)_{\rm MW} = 0.025$ mag and the distance modulus is     $\mu = 39.63 \pm 0.01$.   
Due to the small number of data points, we fit 2nd order polynomial functions to the
resimulated        data. The $M_{\rm V,peak}$ and $\Delta m_{\rm V,15}$ for this system are corrected with the SN 1998bw peak SED and decline rate templates.  
The results are shown in Figure~\ref{fig:sn2003dh} (left). %The upper panel  shows the observational data, the resimulated   data, and the fitting light curves. After correcting the extinction and the distance modulus,   the light curves      in the rest frame V band are shown in the lower panel.  

%%%%%%%%%%%%%%%%%%%%%%%%%%%%%%%%%%%%%%%%%%%%%%%

\subsection{GRB 031203/SN 2003lw}

GRB 031203 (SN 2003lw) 
had a very faint afterglow  and a  relatively bright host galaxy with   ${\rm V}_{\rm host} = 20.57 \pm 0.05$ mag and ${\rm R}_{\rm host} = 20.44 \pm 0.02$ mag \citep{mazzali_2003lw, malesani_talk}. It is a class $A$ system.

The redshift is $z = 0.1055 \pm 0.0001$  %\citep{prochaska_2003_1, 
\citep{prochaska_2004}. We collect    data   in the V and  R bands \citep{malesani_2003lw_one, mazzali_2003lw}. 

 The observed fluxes are corrected for  
 the significant host contribution. After that,  we fit  2nd  order polynomial functions to the resimulated   data.  Unfortunately,  the V band data do not cover the rising part of the light curve. 
 Therefore, only the R band data is used  to   generate the light curves.  
 This system has uncertain extinction.  From \cite{prochaska_2004},  a lower   Galactic extinction is adopted to be $E(B-V)_{\rm MW} =  0.78$ mag, and total extinction is  $E(B-V)_{{\rm total}} =  1.17 \pm 0.1$ mag, through Balmer line ratio study.    Through spectral modeling,    %$E(B-V)_{\rm host} =$ 0.25, 0.3 and 0.35  mag   reproduce Ca$_{\rm II}$-O$_{\rm I}$ absorption well 
  \citep{mazzali_2003lw}   favors a value of the host extinction $E(B-V)_{\rm host} =$ 0.25 mag  and  $A_{\rm V, host} = 0.78 \pm 0.16$  mag  \citep{cardelli_1998}  %$E(B-V)_{\rm host} \sim  0.25 \pm 0.05$ 
  and total reddening $E(B-V)_{\rm total} \sim  1.07 \pm 0.05$ mag.  We adopt the Galactic extinction to be  $E(B-V)_{\rm MW} =  1.06$ mag and the host extinction to be   $A_{\rm V, host} = 0.78 \pm 0.16$ %$E(B-V)_{\rm host} =  0.25  \pm 0.05$ 
 mag from  \cite{mazzali_2003lw}.  We consider the peak magnitudes are uncertain values, because the   host extinction is uncertain.        The distance modulus is    $\mu = 38.52 \pm 0.02$.   
  The $M_{\rm V,peak}$ and $\Delta m_{\rm V,15}$ for this system are corrected with the SN 1998bw peak SED and decline rate templates. 
  The results are shown in Figure~\ref{fig:sn2003dh} (right).

 % However, we consider the peak magnitude somewhat uncertain value because of the possible zero point errors in the photometry and the uncertain extinction correction.

% $E(B-V)_{\rm MW} =  1.04$ mag  from \cite{prochaska_2004}, while  \cite{malesani_2003lw_one} suggested a lower limit as $E(B-V)_{\rm MW} =  0.78$ mag.  The total extinction (both the Galactic and the host) is estimated to be $E(B-V)_{{\rm total}} =  1.07 \pm 0.05$ mag  \citep{mazzali_2003lw}. 

  %%%%%%%%%%%%%%%%%%%%%%%%%%%%%%%%%%%%%%%%%%%%%%%%
\subsection{GRB 050525A/SN 2005nc}

%%%%%%%%%%%%%%%%%%%%%%%%%%%%%%%%%%%%%%%%%%%%%%%
  \begin{figure*}
 \centering
\includegraphics[width=8.8cm]  {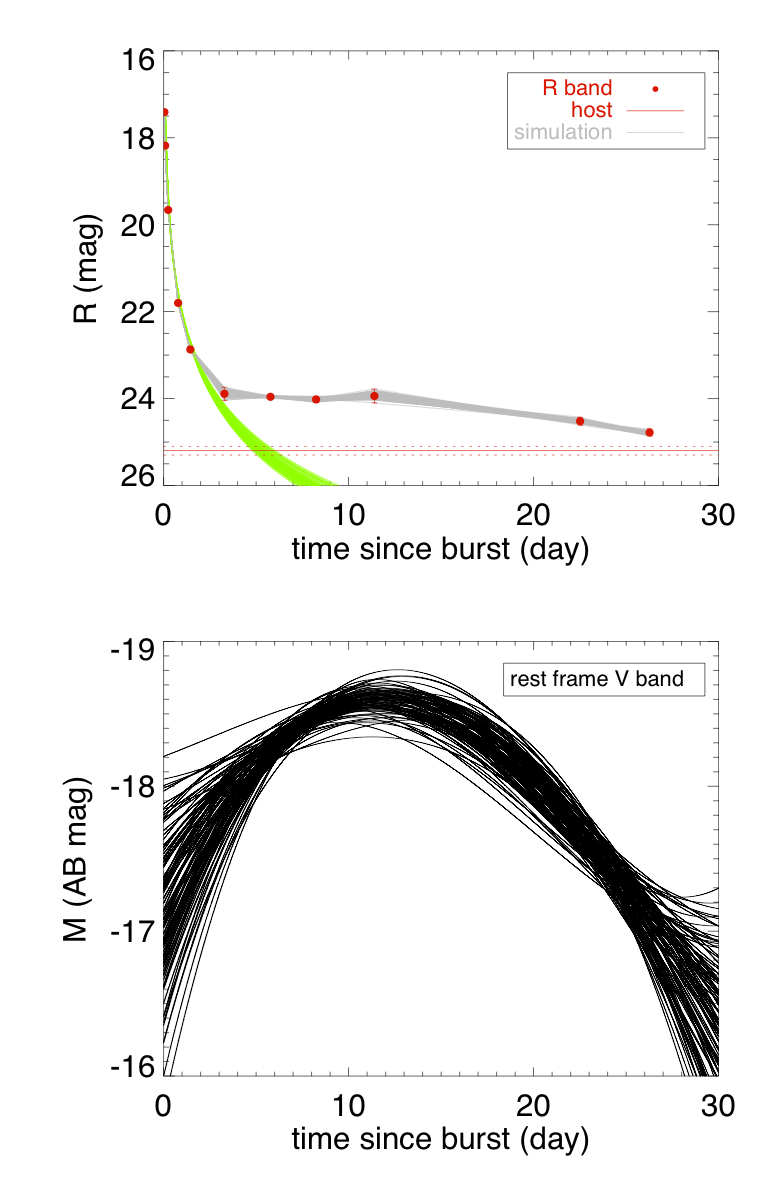}
\includegraphics[width=8.8cm]{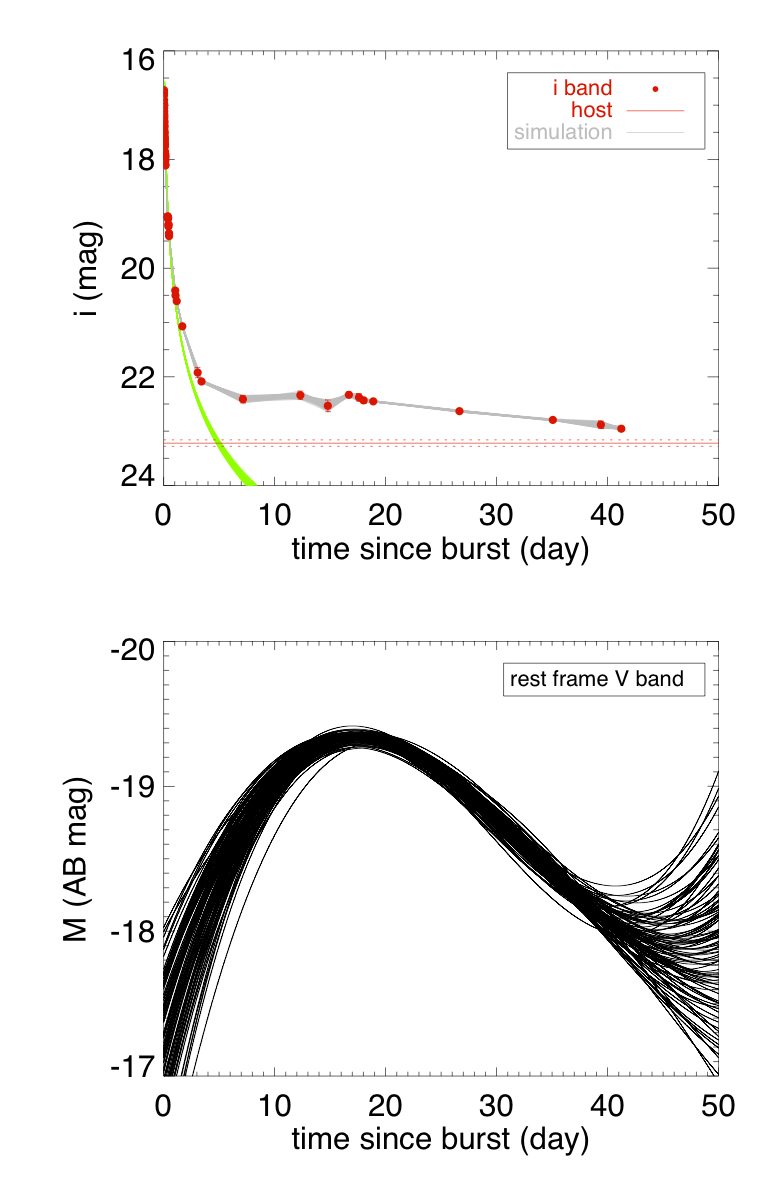}
\caption{GRB 050525A /SN 2005nc ({\it   left}) and GRB 090618  ({\it   right}).  The line styles are    the same  as in Figure \ref{sn1998}.  The green lines are the broken power-law functions fitted to the afterglow.    
 %The   uncertainties in  the resimulated data and the fitting light curves are  plotted  as 68.3\% $ (\pm 1  \sigma) $ of the total resimulated results.  
   }     
 \label{fig:grb050525}
\end{figure*}
%%%%%%%%%%%%%%%%%%%%%%%%%%%%%%%%%%%%%%%%%%%%%%%%
 
GRB 050525A (SN 2005nc) is a     long GRB    with   redshift  $z = 0.606$    \citep{blustin_050525}. % \citep{foley_050525}.   
It is a class $B$ system.
% \citep{gehrels_2005}

We collect   data  from \cite{dellavalle_050525}.   Only R band data is available.   Therefore, the SN 1998bw peak SED and decline rate templates  are applied.  We subtract the host  contribution   with ${\rm R}_{\rm host} = 25.2 \pm 0.1$ mag  \citep{dellavalle_050525}.   %, estimated with the spectrum subtracted from   a template of a blue star-forming galaxy   
  Then we resimulate    and subtract the  afterglow data, fitted as a broken power-law, to get the intrinsic SN  flux.   After that,  we fit  3rd order polynomial functions to the resimulated   data.   The host extinction  is estimated to be $A_{\rm V, host} = 0.26 \pm 0.12$  %$E(B-V)_{\rm host} = 0.09 \pm 0.04$ 
  mag  \citep{cardelli_1998, pei_rv,  blustin_050525}, assuming an  SMC extinction curve.    For the Galactic extinction, %in reference  \citep{blustin_050525}, the value is $A_R = 0.25$ mag.    W
the foreground extinction is $E(B-V)_{\rm MW} = 0.094$ mag. The distance modulus is     $\mu = 42.84 \pm 0.004$.  
Figure  \ref{fig:grb050525} (left) shows the results for SN 2005nc.  %The upper panel shows  the observational data,   the resimulated   data  and the host galaxy magnitude in the R band.     After the subtraction of brightness of  the host and the afterglow contributions, the light curves in   absolute magnitude in the rest frame R   band are presented in the lower panel.   
 %The $M_{\rm V,peak}$ and $\Delta m_{\rm V,15}$ for this system are corrected with SN 1998bw peak SED and decline rate templates. 

%%%%%%%%%%%%%%%%%%%%%%%%%%%%%%%%%%%%%%%%%%%%%%%%

\subsection{XRF 060218/SN 2006aj}
\label{subsec:2006aj}

%%%%%%%%%%%%%%%%%%%%%%%%%%%%%%%%%%%%%%%%%%%%%%%
   \begin{figure*}
 \centering
\includegraphics [width = 17.6 cm]{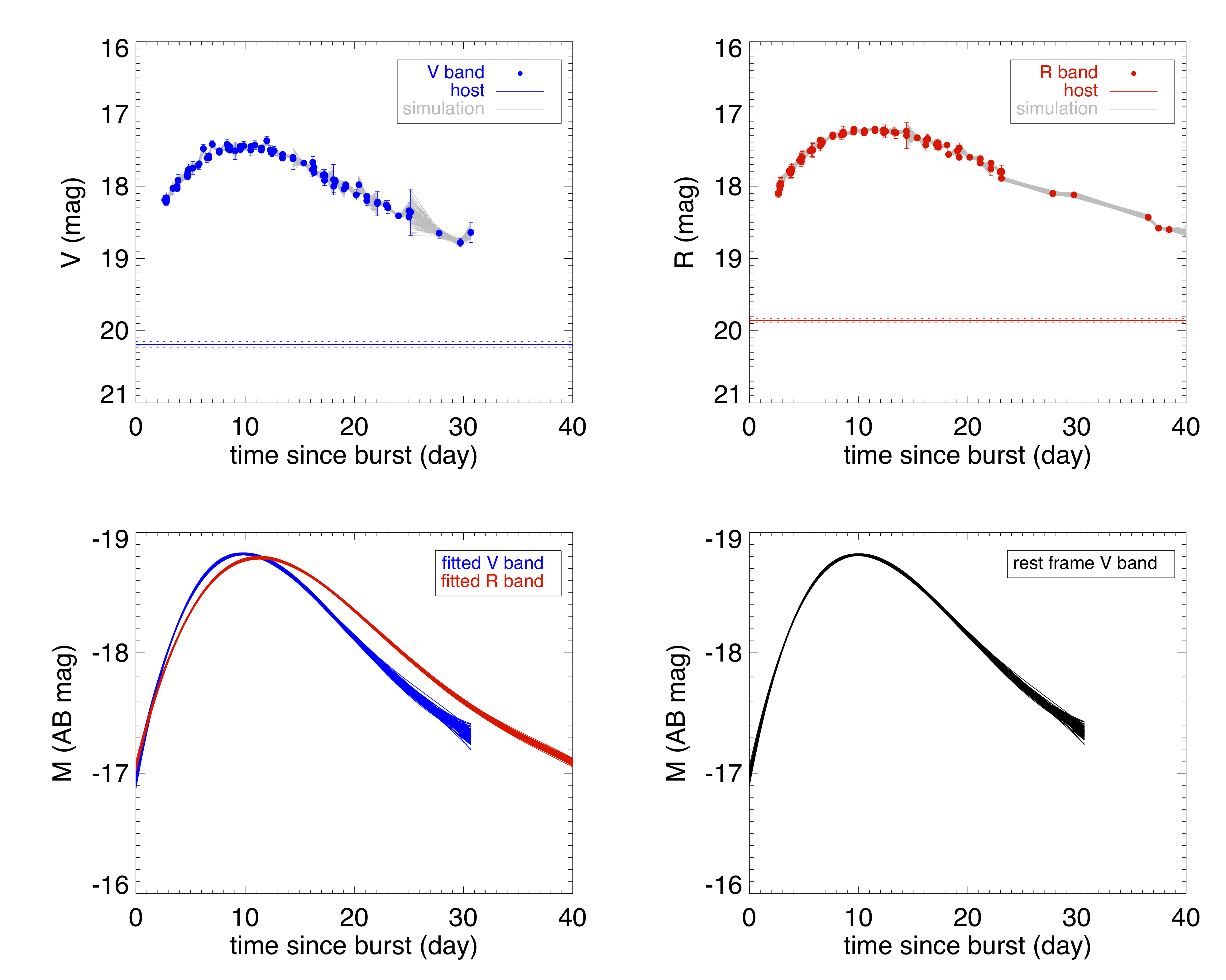}%{sn2006aj.pdf}
\caption{ XRF 060218 (SN 2006aj).    The line styles are  the same  as in Figure \ref{sn1998}.   
 %The   uncertainties in  the resimulated data and the fitting light curves are  plotted  as 68.3\% $ (\pm 1  \sigma) $ of the total resimulated results.     %The upper   panels  shows the observational data, the resimulated   results   and the   host brightness in the V band (left) and    in the R band (right). In the lower left panel, the light curves after the extinction and  distance modulus corrections  are plotted in the V band (blue) and the R band (red). In the lower right panel, the final light curves after   K-correction in the rest frame V band are plotted. 
}
  \label{sn2006aj}
\end{figure*}
%%%%%%%%%%%%%%%%%%%%%%%%%%%%%%%%%%%%%%%%%%%%%%%%

The X-Ray Flash \citep[XRF; ][]{heise_2001}  060218 is a long GRB.    % \citep{cusumano_2006}. %, with relatively low peak energy: $E_{\rm iso} \sim 5 \times 10^{49} $ erg in   the $ 15.5 - 154.8$ kev band  \citep{ferrero_2006aj}.  
%It is a class $A$ system.
The redshift is $z = 0.03342$.   This system is another class $A$ GRB-SN.

We collect data from \citet{sollerman_2006aj}, \citet{ferrero_2006aj}, and \citet{simon_2006aj} in the V and R  bands.    SN 2006aj is an extreme case  \citep{simon_2006aj} because before a normal SN peak, there is an   early peak and these two peaks are  equally  bright.    The data from  \citet{simon_2006aj} includes the early part    ($<$ 2.5 days) of the photometry  since the burst, therefore the light curve shows two bumps. In this paper, we study the normal SN light curve so we only collect the data after 2.5 days since the burst.

  SN 2006aj is located in a relatively bright host galaxy with ${\rm V}_{\rm host} = 20.19 \pm 0.04$  mag and ${\rm R}_{\rm host} = 19.86 \pm 0.03$ mag \citep{sollerman_2006aj}, which are subtracted from the observed fluxes.  The distance modulus is    $\mu = 35.92 \pm 0.07$.     
We fit 4th order polynomial functions.     There is a discrepancy in  the reported estimates of  the host extinction.  \citet{campana_2006} estimate it to be $E(B-V)_{\rm host} = 0.2 \pm 0.03$ mag, assuming  an SMC reddening law.  %thermal radiation detected by VOT.  
%%from an analysis of the Na I D2 absorp- tion lines in our Galaxy (AV = 0.39 mag) and the host galaxy (AV = 0.13 mag) in a VLT UVES spectrum
%From an analysis of the Na I D2 absorption lines and 
Assuming the  relation between sodium absorption and dust extinction from \cite{munari_1997} is representative for interstellar medium,  \cite{guenther_2006} find $E(B-V)_{\rm host} = 0.042 \pm 0.003$ mag.      %and  $E(B-V)_{\rm MW} = 0.127 \pm 0.005$ mag %.  With sodium lines, the total extinction is estimated to be  $E(B-V)_{{\rm total}} = 0.145$ mag by \cite{guenther_2006} 
  With an updated empirical relation from \cite{poznanski_extinction}, the extinctions are %$E(B-V)_{\rm host} = 0.027_{-0.009}^{+0.014}$ mag    and  $E(B-V)_{\rm MW} = 0.061_{-0.019}^{+0.029}$ mag, which are about half of the 
$E(B-V)_{\rm host} = 0.026 \pm 0.014$ mag    and  $E(B-V)_{\rm MW} = 0.061 \pm 0.03$ mag, which are about half of the 
values from  \cite{munari_1997}.  
 We adopt  the Galactic extinction  is    $E(B-V)_{{\rm MW}} = 0.145$ mag.  The host extinction is estimated to be  $E(B-V)_{\rm host} = 0.026 \pm 0.014$ mag  and $A_{\rm V, host} = 0.076 \pm  0.041$ mag  with the updated empirical relation from  \cite{poznanski_extinction}.   We apply the multi-band K-correction with $ c  =  0.87$.   %{guenther_2006, poznanski_extinction}. 
%We derive an estimate by \citep{sollerman_2006aj} with $E(B-V)_{{\rm total}} = 0.145$ mag, which is   consistent with the total extinction by \cite{munari_1997}, using the sodium lines.     
% The   peak magnitude is  constrained to be $M_{\rm V,peak} =   -18.77^{+0.07}_{-0.07}$ mag and the decline rate in 15 days is $\Delta m_{\rm V,15} = 1.08^{+0.18}_{-0.13}$ mag.     The results are consistent with the   results of $M_{\rm V,peak} = -18.7$ mag  and  $\Delta m_{\rm V,15} = 1.1 \pm 0.1$ mag   from \citet{sollerman_2006aj}. 
Figure~\ref{sn2006aj} shows   the light curves of SN 2006aj.   %In the upper panel, the observed and resimulated   data and the host galaxy brightness    in V band (left) and R band (right) are plotted.       In the lower panel,  the left plot  shows the light curves in the V and R bands, after correcting for Galactic extinction and converting into the absolute magnitude.  The lower right panel shows     the   resulting light curves in the rest frame V band, after multi-band K-correction. 

%_{-0.009}^{+0.014}

 %%%%%%%%%%%%%%%%%%%%%%%%%%%%%%%%%%%%%%%%%%%%%%%%
\subsection{GRB 090618}

 %%%%%%%%%%%%%%%%%%%%%%%%%%%%%%%%%%%%%%%%%%%%%%%

The long GRB 090618   %\citep{schady_2009}  %, cenko_2009, katkhullin_2009}  
    is a class $C$ system with  $z =  0.54$ \citep{cano_060729}. We collect   data  from \cite{cano_060729} in the    i band.      We subtract the brightness of the host galaxy and the afterglow. The   host brightness   is estimated to be  $i_{\rm host} = 23.22 \pm 0.06$ mag \citep{cano_060729}. 
    The afterglow is fitted with   broken power-law functions and the resimulated        data are fitted with 3rd order polynomial functions.  The Galactic extinction  is  $E(B-V)_{\rm MW} = 0.09$ mag.  From X-ray to optical SED fitting,   the host extinction   is $A_{\rm V, host} = 0.3 \pm 0.1$ mag according to \cite{cano_060729}.   
The distance modulus is    $\mu = 42.54 \pm 0.004$.    
  The SN 1998bw peak SED and decline rate templates  are used to convert the peak magnitude and the decline rates into the rest frame V band.  
 Figure~\ref{fig:grb050525} (right) shows     the  results for GRB 090618.    % In the upper panel, the observational  data, the resimulated   data, the host brightness  in R band (left) and I band (right) are plotted.    In the lower panel, on the left,  the absolute magnitude in the R and I bands are plotted, after correcting for the host and afterglow contributions, Galactic extinction, as well as the distance modulus. In the lower right panel, after multi-band   K-correction, the light curves in the rest frame V band are plotted.  

%$R_{\rm host} = 23.44 \pm 0.06$  mag and

%%%%%%%%%%%%%%%%%%%%%%%%%%%%%%%%%%%%%%%%%%%%%%%%
\subsection{XRF 100316D/SN 2010bh}
\label{sub:2010}

%%%%%%%%%%%%%%%%%%%%%%%%%%%%%%%%%%%%%%%%%%%%%%%
   \begin{figure*}
 \centering
\includegraphics [width = 17.6 cm]{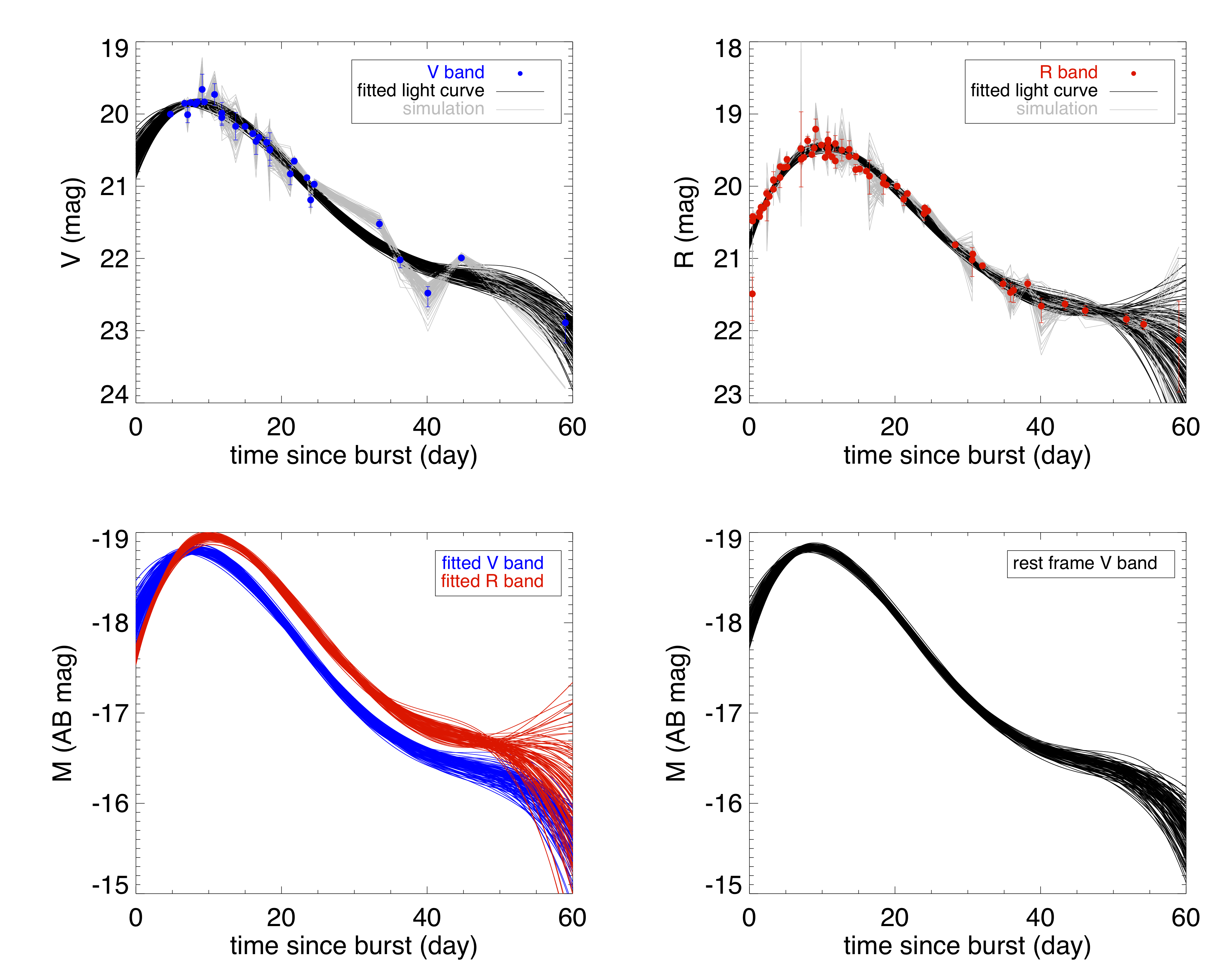}%{sn2010bh.pdf}
\caption{ XRF 100316D/SN 2010bh.     The line styles are  the same  as in Figure \ref{sn1998}.   
 %The   uncertainties in  the resimulated data and the fitting light curves are  plotted  as 68.3\% $ (\pm 1  \sigma) $ of the total resimulated results.    % The upper left panel  shows the observational data, the resimulated   results  and the   polynomial functions fitted to the resimulated   data in the V band. The upper right panel  shows the corresponding results   in the R band. In the lower left panel, the light curves after the extinction and the distance modulus correction are plotted in the V (blue) and R (red) bands. In the lower right panel, the final light curves after the K-correction in the rest frame V band are plotted.  
 }
  \label{sn2010bh}
\end{figure*}
%%%%%%%%%%%%%%%%%%%%%%%%%%%%%%%%%%%%%%%%%%%%%%%%

XRF 100316D is a soft long GRB   \citep{cano_2010bh}.    % \citep{vergani_2010}. 
 It is a class $A$ system. The redshift is $z = 0.059$ and we use published photometry in  the V  and  R  bands  \citep{cano_2010bh, olivares_e_2010bh, bufano_2010bh}.   The three data sets   are not consistent with each other (shown in Figure~\ref{sn2010bh}).  There are systematic offsets in the photometry,   especially around the peak. Compared to        \cite{bufano_2010bh} and \cite{ cano_2010bh}, the R band data from \cite{olivares_e_2010bh}  is about  0.3 mag fainter  at the peak.   This may because of zero point discrepancies. We reduce the offset by subtracting  0.3 mag from the R band data   \citep{olivares_e_2010bh}, although we acknowledge there is a possibility that the other two data sets should be shifted instead. %It is dangerous to offset some datasets to the others, without knowing how exactly the data come from. So we keep the original data without shift any of the data set  to the other one. 

The foreground extinction is $E(B-V)_{\rm MW} = 0.117$ mag.  
%The value is consistent with the values estimated %by  \citet{bufano_2010bh},  using the strength of
%%the Na I D absorption lines: $E(B-V)_{\rm MW} = 0.12$ mag and  
%by \cite{olivares_e_2010bh}:  $E(B-V)_{\rm MW} = 0.117$ mag.
 Reported values of the  host extinction are very different.  Using the H$\alpha$/H$\beta$ ratio, the host extinction is estimated to be $E(B-V)_{\rm host} = 0.14$ mag \citep{bufano_2010bh}. 
 From color excess measurement,   \cite{cano_2010bh} assumes the host  extinction to be $E(B-V)_{\rm host} = 0.18 \pm 0.08$ mag.    
 %was inferred by  \cite{cano_2010bh}, which implies   $E(B-V)_{\rm host}   \sim 0.06$ mag.
  \cite{olivares_e_2010bh} estimated the extinction using afterglow SED fitting and found 
$A_{\rm V, host} = 1.20 \pm 0.09$ mag.  %$E(B-V)_{\rm host} = 0.39 \pm 0.03$ mag. %, which is consistent at the 2.5$\sigma$ confidence level with the value reported by \citet{cano_2010bh}. 
We adopt this value      %$E(B-V)_{\rm host} = 0.39 \pm 0.03$ mag 
because the intrinsic SN spectrum is otherwise very red (\cite{levan_130427}).    However, we consider the peak magnitude     an uncertain value because of the possible zero point errors in the photometry and the uncertain extinction correction.
%Thus, the   peak magnitude is $M_{\rm V,peak} = -18.73^{+0.06}_{-0.06}$ mag, which is   consistent with the value $M_{\rm V,peak} = -18.62 \pm 0.08$ mag estimated from  \cite{cano_2010bh}.  

     The distance modulus is   $\mu = 37.20 \pm 0.04$.    
 We fit 4th order polynomial functions to the resimulated   data. The multi-band K correction parameter is $c =  0.77$. 
 Figure~\ref{sn2010bh}   shows the resulting light curves for SN 2010bh.   % In the upper panel, the observational data, the resimulated   data and the fitting functions to the resimulated   data in V band (left) and R band (right) are plotted. 
 %  The lower panel  on the left  shows the light curves in V and R band, after correcting the Galactic extinction   and the host extinction, as well as the distance modulus.  On the lower right panel, after multi-band K-correction, the  light curves in the rest frame V band are plotted.    

%%%%%%%%%%%%%%%%%%%%%%%%%%%%%%%%%%%%%%%%%%%%%%%%
\subsection{GRB 120422A/SN 2012bz}

%%%%%%%%%%%%%%%%%%%%%%%%%%%%%%%%%%%%%%%%%%%%%%%
   \begin{figure*}
 \centering
\includegraphics [width = 17.6 cm] {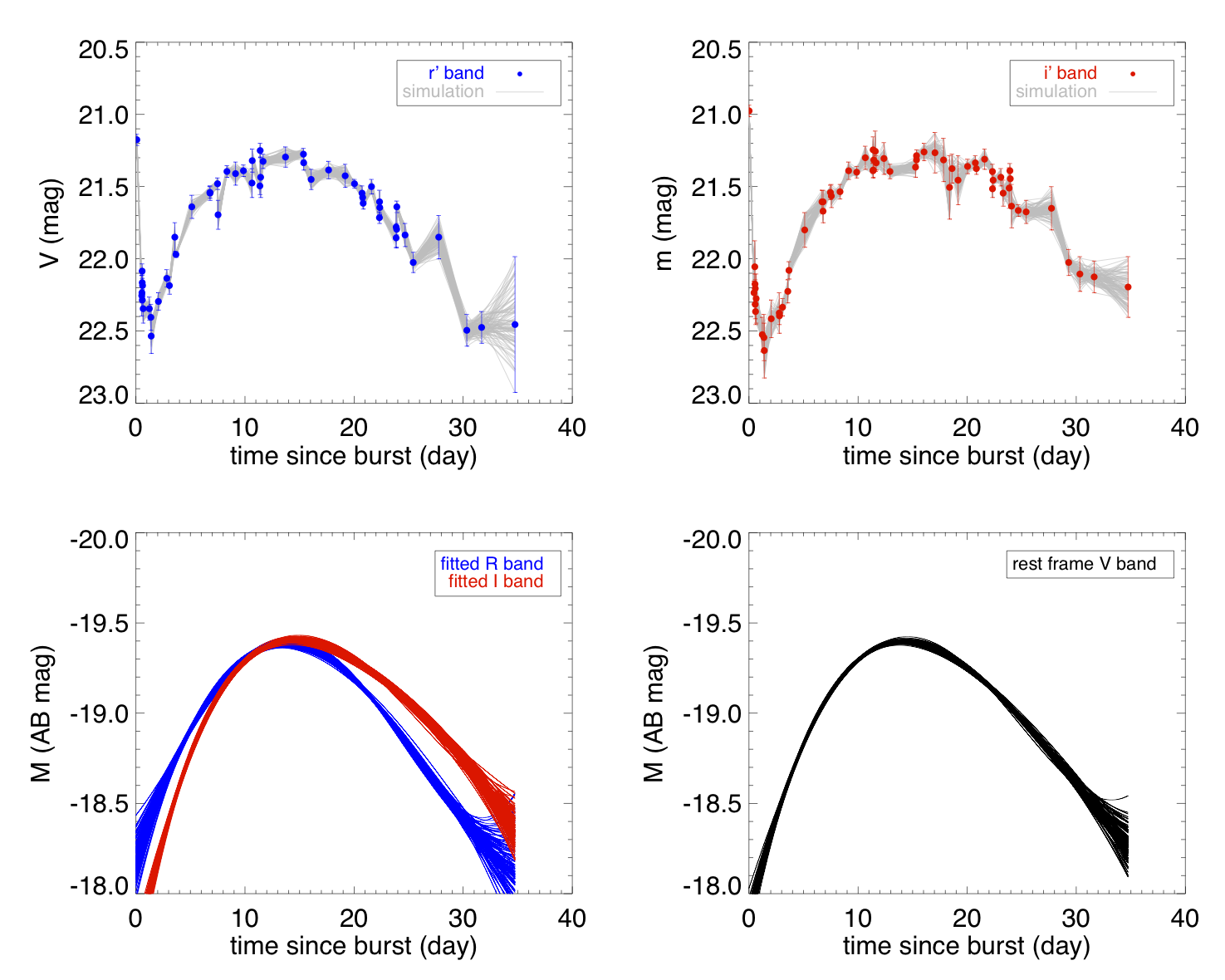}%{sn20112bz.pdf}
\caption{ GRB 120422A/SN 2012bz.      The line styles are  the same  as in Figure \ref{sn1998}.  
 %The   uncertainties in  the resimulated data and the fitting light curves are  plotted  as 68.3\% $ (\pm 1  \sigma) $ of the total resimulated results.    %The upper left panel  shows the observational data, the resimulated   results  and the   polynomial functions fitted to the resimulated   data in R band. The upper right panel shows the similar results   in I band. In the lower left panel, the magnitudes are transferred into the absolute magnitude and the light curves  in R and I bands are plotted.  In the lower right panel, after the K-correction, the light curves in the rest frame V band are plotted. 
}
  \label{sn2012bz}
\end{figure*}
%%%%%%%%%%%%%%%%%%%%%%%%%%%%%%%%%%%%%%%%%%%%%%%%

   Extensive observations have been done to detect GRB 120422A (SN 2012bz)   with   telescopes from mm to optical wavelengths   \citep{melandri_2012bz, schulze_2012bz}.  It is a class $A$ system.

The redshift   is $z = 0.283$, and the data are collected from \cite{melandri_2012bz, schulze_2012bz}  in the r'  and i' bands.  Compared to the X-ray lightcurve (Fig.\ 2 in \cite{schulze_2012bz}), the afterglow  in   the r' and i' bands have a significant supernova contribution, so  we fix the  the  post-break slope $\beta = 1.48 \pm 0.4$ based on the X-ray observations \citep{schulze_2012bz}.  But  the subtraction of the afterglow barely changes the intrinsic SN brightness. So for this system, either we fix the slope based on X-ray observation or on the   SN modeling have no difference to  the brightness of the  SN  \citep{schulze_2012bz}. 
% The slope of the post-break decay is fixed to be $\beta_2 = 2$ (Eq. \ref{afterdecline}), based on the SN modelling      \citep{schulze_2012bz}.  
 We adopt the   foreground extinction to be $E(B-V)_{\rm MW} = 0.035$ mag.  The resimulated  data are fitted with 4th order polynomial functions. 
    The distance modulus is      $\mu = 40.89 \pm 0.01$.   The   multi-band K-correction parameter  is $c = 0.40$. 
 Figure \ref{sn2012bz}  shows the light curve results.   %The upper panels show  the observational data, the resimulated   data and the fitting functions to the resimulated   data.    The lower panels  show the light curves in R and I band (left), and  the   light curves in the rest frame V band (right).   

%%%%%%%%%%%%%%%%%%%%%%%%%%%%%%%%%%%%%%%%%%%%%%%%

\subsection{GRBs not included}
\label{noincludegrb}
We   investigated the possibilities of obtaining light curves for other   GRB-SNe in  class $A$, $B$ and $C$.  Here we briefly explain the   reasons why we do not report the light curves for     these systems.

  There are several reasons that may cause the failure of obtaining the light curves in the rest frame V band: 1) 
  The errors in the subtraction of the afterglow will inflate the errors in the SN photometry. This is a major reason why for some systems, even though   enough data points have been obtained, after the afterglow fitting, there are too few useful data points left to do  the polynomial fitting. We cannot get   full light curves (notably information at and before the peak) for these systems.  2) 
Some systems    lack data in   proper band(s)       to do the      K-correction.  This is because the multi-band K-correction is only valid when the redshifted V band is between the two observed bands. 3) Some systems have very uncertain host extinction or host galaxy contribution.  4) Some systems lack enough data to do polynomial fits to obtain light curves. e.g., for   2nd order, at least 3 data points are required.  In practice, to obtain well defined light curves, more data points are needed. 
Below we provide a brief discussion for each system.   The reasons for these systems being excluded from our analysis are summarized in Table \ref{unable}. 

%%%%%%%%%%%%%%%%%%%%%%%%%%%%%%%%%%%%%%%%%%%%%%%%

\begin{table}[h!tbp]
\begin{center}
\caption{A list of unselected systems in class $A$, $B$, and $C$.   }
 \label{unable}
 \begin{tabular}{cc} 
    \hline
 \hline
   GRB/XRF/SN &   reason(s) \\
    \hline
 970228 &  4 \\
   990712 &  1, 4 \\
  011121/2001ke & 3, 4 \\
 020903 & 1 \\
   021211/2002lt &   1 \\
   041106 & 2 \\
  080319B & 1 \\
  081007/2008hw & 3\\
 091127 & 1 \\
  101219B/2012ma & 4 \\
120714B/2012eb & 4 \\
  
    \hline
% \hline
                                                                   
\end{tabular}
\end{center}
{\footnotesize
 \noindent
1: After subtracting the afterglow, there are too few data points left to obtain full light curves.  

\noindent
2: Lack of data in proper band(s) to do K-correction. 

\noindent
3: Uncertain host extinction or host galaxy contribution. 

\noindent
4: Lack of sufficient data around the peak to do polynomial fitting. 
}
\end{table}

 %%%%%%%%%%%%%%%%%%%%%%%%%%%%%%%%%%%%%%%%%%%%%%%%

\begin{description}

  \item[\textbf{GRB 970228}] At $z = 0.695 \pm 0.002$    \citep {galama_970228}   I band data are collected. However, there are not enough  data ($<$ 3) around the peak in   the I band. 
  \item[\textbf{GRB 990712}]   The V and I bands    \citep{bjrnsson_990712, christensen_990712, sahu_990712} have less than  3 data points after   subtracting the afterglow brightness.  In   the R band, there are not enough  data   points to perform     polynomial fitting. 
   \item[\textbf{GRB 011121/SN 2001ke}] At  $z = 0.36$ \citep{bloom_011121}    R and I band data   \citep{bloom_011121, garnavich_011121, greiner_011121, kupcuyolda_011121} are collected. The host   and the afterglow brightness are subtracted. Then there are not enough data around peak to do polynomial fitting in the I band. In addition, the host extinction is  very  uncertain.
  \item[\textbf{XRF 020903}] After subtracting the afterglow brightness, there are not enough useful data points to do the polynomial fitting \citep{bersier_020903, soderberg_020903}. 
  \item[\textbf{GRB 021211/SN 2002lt} ] After subtracting the host and afterglow flux, there are too few data points left ($<$ 3) \citep{dellavalle_2001lt} to obtain     light curves.  
  \item[\textbf{GRB 041006}] Only in the R band there  are enough data \citep{stanek_041006} to extract the light curve. But at a redshift of $z = 0.716$,  the R band is too far from the rest frame V band.

 \item[\textbf{GRB 080319B} ]

The redshift is $z = 0.937$  and  the data \citep{tanvir_080319b, bloom_080319b} are in the R and I bands.  The host contributions are $R_{\rm host} = 26.96 \pm 0.13$ mag and $I_{\rm host} = 26.17 \pm 0.15$ mag.   The  afterglow slope is fixed to $\beta_2 = 2.33$ \citep{tanvir_080319b, bloom_080319b}.   After subtracting the afterglow brightness there are   not enough useful data points left to fit   polynomial functions.

    \item[\textbf{GRB 081007/SN 2008hw} ] The redshift is $z = 0.5295$  and the data  are in the r' and i' bands \citep{jin_2008hw}.  
For the afterglow fitting, we fixed the slope to $\beta_2 = 1.25$,   based on   the X-ray observation  \citep{jin_2008hw}.  A multi-band K-correction is applied.   But the host contribution is uncertain. If we assume  it has host $r_{\rm host} = 25.0$ mag and $i_{\rm host} = 24.5$ mag, then the peak magnitude is $M_{\rm V,peak} = -18.85^{+0.91}_{-0.64}$ mag.  However, a different estimate of the host galaxy brightness would lead to different peak magnitudes.

 \item[\textbf{GRB 091127} ]

The redshift is $z = 0.49$ and the data are collected from \cite{troja_091127, vergani_091127, cobb_091127}. 
The i band data are selected.
We subtract the host brightness with $I_{\rm host} = 22.54 \pm 0.10$ mag \citep{troja_091127}. The afterglow is fitted with   broken power-law functions. But after the afterglow fitting, there are not enough data to obtain     light curves and measure the peak magnitude and the decline rate.  %The peak magnitude may be estimated to be $M_{\rm V,peak} = -20.12^{+3.26}_{-1.21}$. 

  \item[\textbf{GRB 101219B/SN 2010ma} ]  With only two data points 
  and an upper limit \citep{sparre_2001ma}, it is not possible to obtain the light curve. 
   \item[\textbf{GRB 120714B/SN 2012eb} ]  %The burst was detected by ${\it   Swift}$  \cite{saxton_120714b}. 
   %GROND/FORS2 
 %  \cite{klose_120714b} detected   a SN.   
    SN 2012eb was confirmed to be associated with GRB 120714B by \cite{grb120714b}.  But there are no published data for SN 2012eb yet. 
   \end{description}

%%%%%%%%%%%%%%%%%%%%%%%%%%%%%%%%%%%%%%%%%%%%%%%%
%%%%%%%%%%%%%%%%%%%%%%%%%%%%%%%%%%%%%%%%%%%%%%%%

\section{Properties of the light curves}
 \label{sec:property}

%%%%%%%%%%%%%%%%%%%%%%%%%%%%%%%%%%%%%%%%%%%%%%%%

\begin{table}[h!tbp]
\begin{center}
\caption{The selected systems and   relevant results     with $ 1  \sigma$ uncertainties.  
}
 \label{property}
 \begin{tabular}{ccccccc} 
  
%\begin{tabular}    {cccccccc@{}}
\hline       
\hline   
 % GRB/XRF/SN &    $z$ &  $M_{\rm V,peak}$      &  $\Delta m_{\rm V,15}$  & $t_{\rm peak}$ &  $ E(B-V)_{\rm MW}$    & $E(B-V)_{\rm host}$        \\ %\tnote{1} \\ 
  GRB/XRF/SN &    $z$ &  $M_{\rm V,peak}^{a}$      &  $\Delta m_{\rm V,15}$  & $t_{\rm peak}$        \\ %\tnote{1} \\ 
                             &             &      (mag)                                &  (mag)                 & (day)                       \\
   \hline

980425/1998bw       & 0.0085 & $-19.29^{+0.08}_{-0.08}$ & $0.75 ^{+0.02}_{-0.02}$   &$16.09^{+0.17}_{-0.18}$     \\
%   && &&\\
%  \hline
  030329/2003dh & 0.1685& $-19.39^{+0.14}_{-0.12}$ & $0.90^{+0.50}_{-0.50}$  & $10.74^{+2.57}_{-0.85}$    \\
%\hline
     031203/2003lw  &  0.1055 & $-19.90^{+0.16}_{-0.16}$ & $0.64^{+0.10}_{-0.10}$  & $19.94^{+1.37}_{-1.48}$       \\
%   \hline
  050525A/2005nc & 0.606  & $-18.59^{+0.31}_{-0.25}$ & $1.17^{+0.69}_{-0.88}$& $11.08^{+2.26}_{-3.37}$       \\
%  \hline

 060218/2006aj     & 0.03342 & $-18.85^{+0.08}_{-0.08}$ & $1.08 ^{+0.06}_{-0.06}$ & $9.96^{+0.18}_{-0.18}$    \\
%  \hline
  090618 & 0.54 & $-19.34^{+0.13}_{-0.13}$ & $0.65^{+0.15}_{-0.17}$  & $17.54^{+1.51}_{-1.64} $   \\
%    \hline

 100316D/2010bh& 0.059  & $-18.89^{+0.10}_{-0.10}$ & $1.10^{+0.05}_{-0.05}$ & $8.76^{+0.31}_{-0.37}$   \\
%   \hline
 120422A/2012bz & 0.283   & $-19.50^{+0.03}_{-0.03}$ & $0.73^{+0.06}_{-0.06}$ & $14.20^{+0.34}_{-0.34}$   \\
%    \hline
 % 020903 &  0.251  & $-18.776^{+0.039}_{-0.093}$ & $1.077^{+0.074}_{-0.064}$ & 0.032  &  -   \\
%  \hline

  \hline
 %\hline
                                                                   
\end{tabular}
\end{center}
{\footnotesize
 \noindent
$a$:   The uncertainties in  $M_{\rm V, peak}$ quadratically come from the polynomial fits, the  $0.02$ mag in K correction, the distance modulus uncertainties and the uncertainties      in the host extinction.   
}
\end{table}

 %%%%%%%%%%%%%%%%%%%%%%%%%%%%%%%%%%%%%%%%%%%%%%%%%

 %%%%%%%%%%%%%%%%%%%%%%%%%%%%%%%%%%%%%%%%%%%%%%%%
   \begin{figure}
 %\centering
\includegraphics [width = 8.8 cm]  {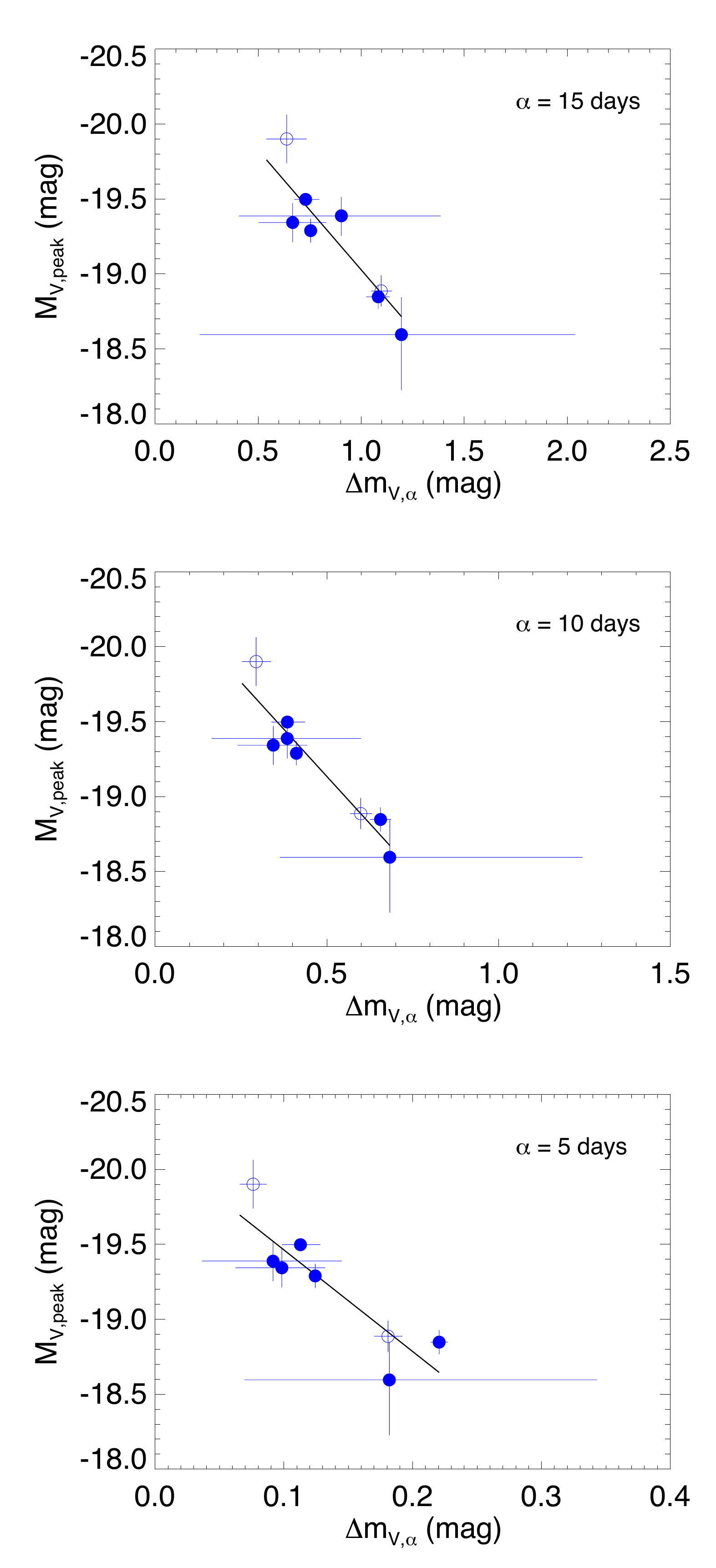}
\caption{  
   The peak-luminosity and decline rate relation for GRB-SNe with the   decline times of     5, 10 and 15 days.  Systems GRB 031203/SN 2003lw and  GRB 100316D/SN 2010bh have uncertain extinction, and they are plotted as open symbols. The best linear fits  to the relations are in       black.    
     }
  \label{relationplotall}
\end{figure}

 %%%%%%%%%%%%%%%%%%%%%%%%%%%%%%%%%%%%%%%%%%%%%%%%
%   \begin{figure}
 %\centering
%\includegraphics [width = 8.8 cm]  {chisquare.pdf}
%\caption{   $\chi^2$ of the   peak-luminosity and decline rate relations   with the decline time  from 3 to  15 days.  
%     }
%  \label{chisquare}
%\end{figure}
 
 %%%%%%%%%%%%%%%%%%%%%%%%%%%%%%%%%%%%%%%%%%%%%%%%

 %%%%%%%%%%%%%%%%%%%%%%%%%%%%%%%%%%%%%%%%%%%%%%%%
   \begin{figure}
 %\centering
\includegraphics [width = 8.8 cm]  {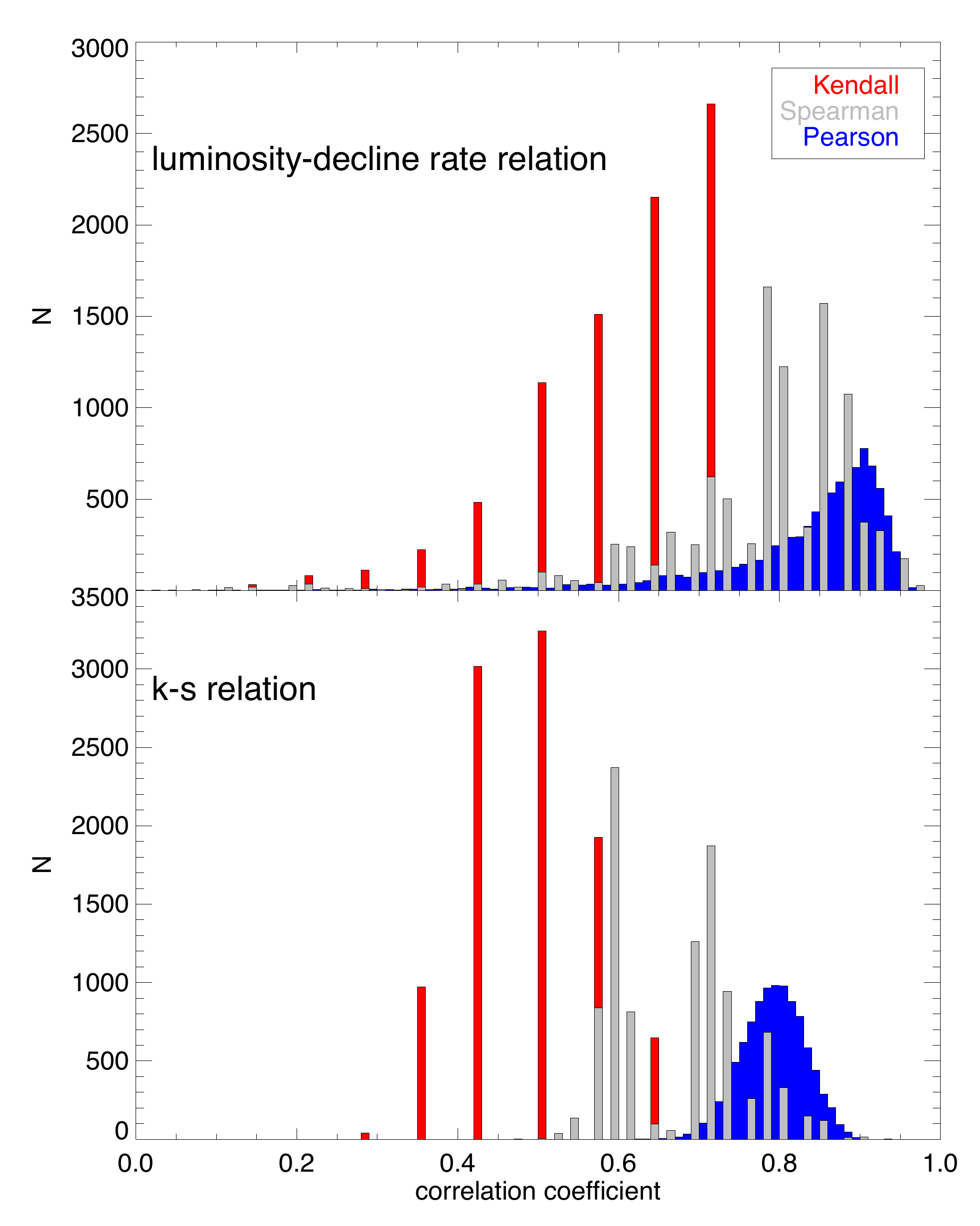}
\caption{  Kendal's $\tau$, Spearman's rank, and   Pearson's correlation coefficients   of %the peak magnitudes and the decline rates \textcolor{red}{at $\alpha = 15$ days}. 
the luminosity-decline rate relation  in the upper panel and the $k-s$ relation  (section \ref{sub:sfactor}) in the bottom panel. Bin size is 0.01. 
 $M_{\rm V,peak}$,     $\Delta m_{\rm V,15}$,  the  $k$ factor and  the $s$ factor are resimulated  10 000 times with the standard Monte Carlo method. 
     }
  \label{coefficient}
\end{figure}
 
 %%%%%%%%%%%%%%%%%%%%%%%%%%%%%%%%%%%%%%%%%%%%%%%%
  %%%%%%%%%%%%%%%%%%%%%%%%%%%%%%%%%%%%%%%%%%%%%%%%
   \begin{figure}
 %\centering
\includegraphics [width = 8.7 cm]  {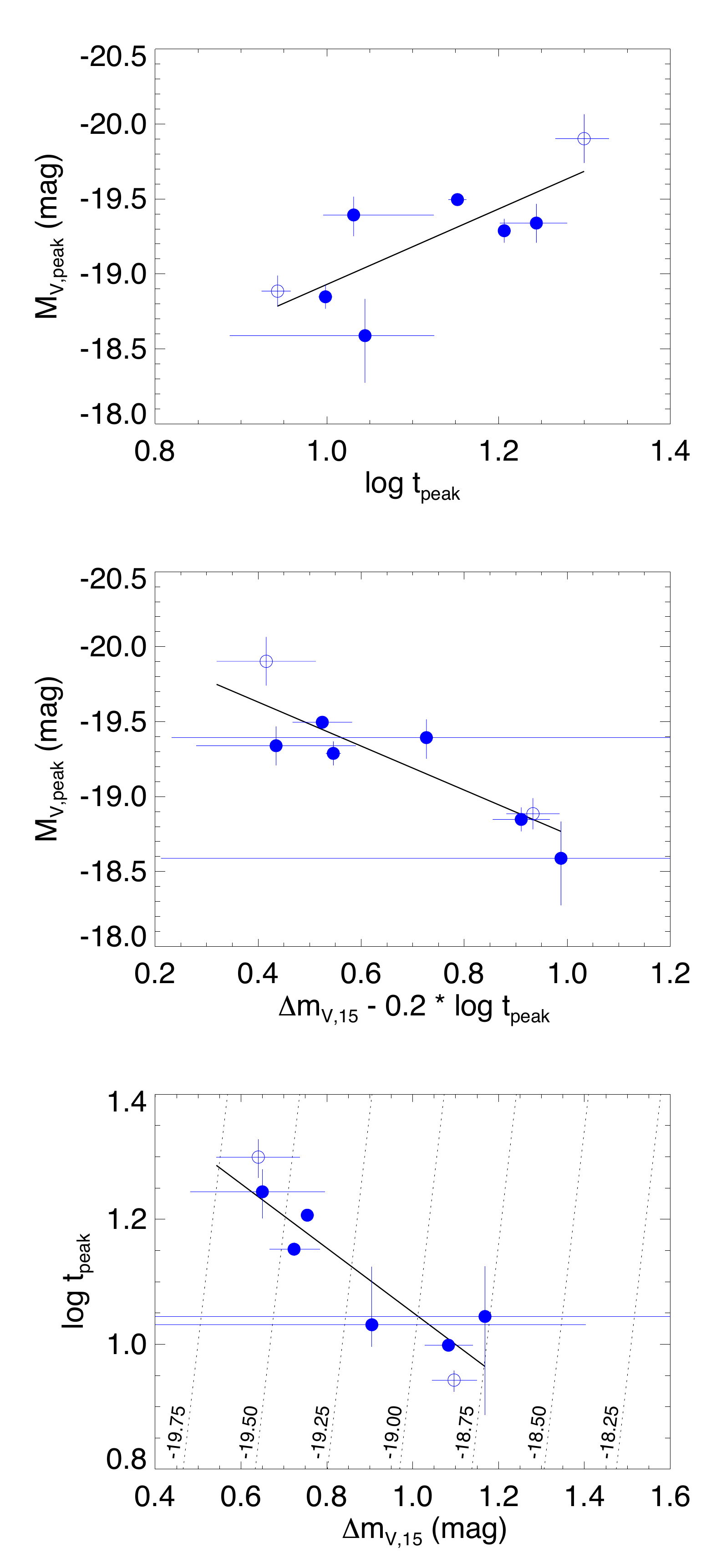}
\caption{  The relations between logarithmic peak time   $\log t_{\rm peak}$,   peak magnitude  $M_{\rm V,peak}$ and    decline rate  $\Delta m_{\rm V,15}$.    The upper panel shows $\log t_{\rm peak}$ as a function of   $M_{\rm V,peak}$.     The middle panel shows the multiple linear relation between   $ \Delta m_{\rm V,15}$,  $\log t_{\rm peak}$ and  $M_{\rm V,peak}$. %,  %as constrained in function \ref{combinetm15}, 
%using least-squares fits. 
The bottom panel displays a `fundamental plane' like relation for GRB-SNe through $ t_{\rm peak}$ and $\Delta m_{\rm V,15}$. The straight lines in blue are the best linear fitting functions (Eqs. \ref{peaklogt}, \ref{m15logt}  and \ref{mvm15logt}). The dotted lines indicate loci of equal absolute peak magnitudes.   The best linear fits are plotted in black.   }
  \label{logtpeak}
\end{figure}
%%%%%%%%%%%%%%%%%%%%%%%%%%%%%%%%%%%%%%%%%%%%%%%%

%   \begin{figure}
% \centering
%\includegraphics [width = 8.8 cm]  {lightcurveall1.pdf}
%\caption{  Shift 1 of light curves of the selected systems in rest frame V band. The magnitude has been shifted as: $M_V ^{'} = M_V \times \Delta \bar  m_{15} / \Delta  m_{15})$, with $ \Delta \bar m_{15}$ representing decline rate of SN 1998bw. Here  $t_{\rm peak}$  and $M_{\rm V,peak}$ have been shifted according to SN 1998bw. Systems GRB031203/SN2003lw and  GRB100316D/SN2010bh have uncertain extinction, and they are plotted in open symbol.  }
%  \label{lightcurveall1}
%\end{figure}
%%%%%%%%%%%%%%%%%%%%%%%%%%%%%%%%%%%%%%%%%%%%%%%%

 %%%%%%%%%%%%%%%%%%%%%%%%%%%%%%%%%%%%%%%%%%%%%%%%
%   \begin{figure}
% \centering
% \includegraphics [width = 8.8 cm]  {lightcurveall2.pdf}
%\caption{  Shift 2 of light curves of the selected systems in rest frame V band. The time has been shifted as: $t^{'} = t \times \bar t_{\rm peak}/t_{\rm peak}$, with $ \bar t_{\rm peak} $ %representing the peak times of SN 1998bw. Here   $M_{\rm V,peak}$ have been shifted according to SN 1998bw.  }
%  \label{lightcurveall2}

%\end{figure}

 %%%%%%%%%%%%%%%%%%%%%%%%%%%%%%%%%%%%%%%%%%%%%%%%
   \begin{figure}
 %\centering
\includegraphics [width = 8.8 cm]    {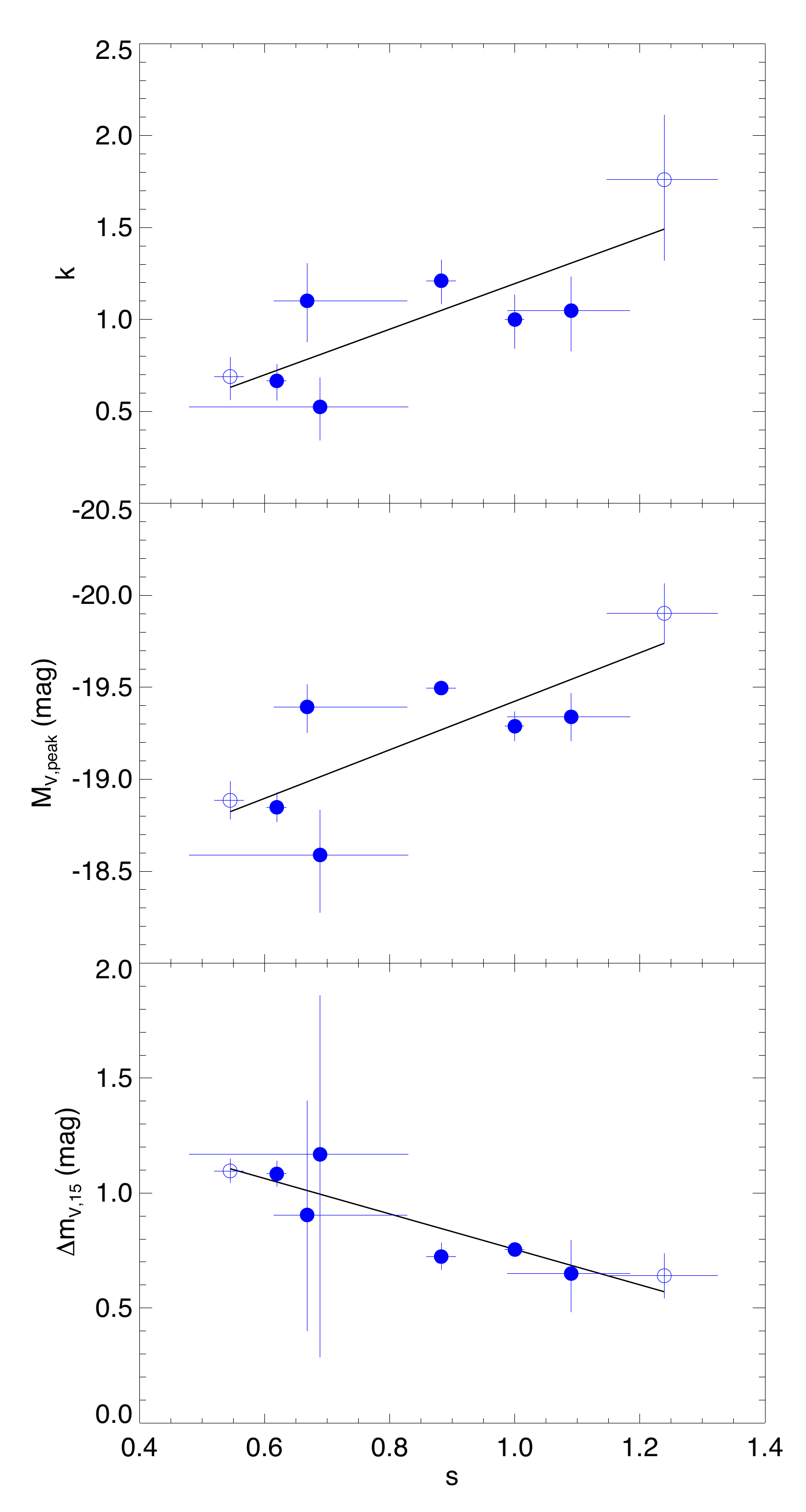}
\caption{   The relations between   the    $s$ factor,     the  $k$  factor,  the  decline rate $\Delta m_{\rm V,15}$ and the  peak magnitude $M_{\rm V,peak}$.    The best linear  fits are  plotted in black. 
%straight line in the upper panel represents the best fitting function to  the relation between $\Delta m_{\rm V,15}$ and $s$. 
 }
  \label{ksfactor}
\end{figure}
%%%%%%%%%%%%%%%%%%%%%%%%%%%%%%%%%%%%%%%%%%%%%%%%

%%%%%%%%%%%%%%%%%%%%%%%%%%%%%%%%%%%%%%%%%%%%%%%%

   \begin{figure}
 \centering
\includegraphics [width = 8.8 cm]  {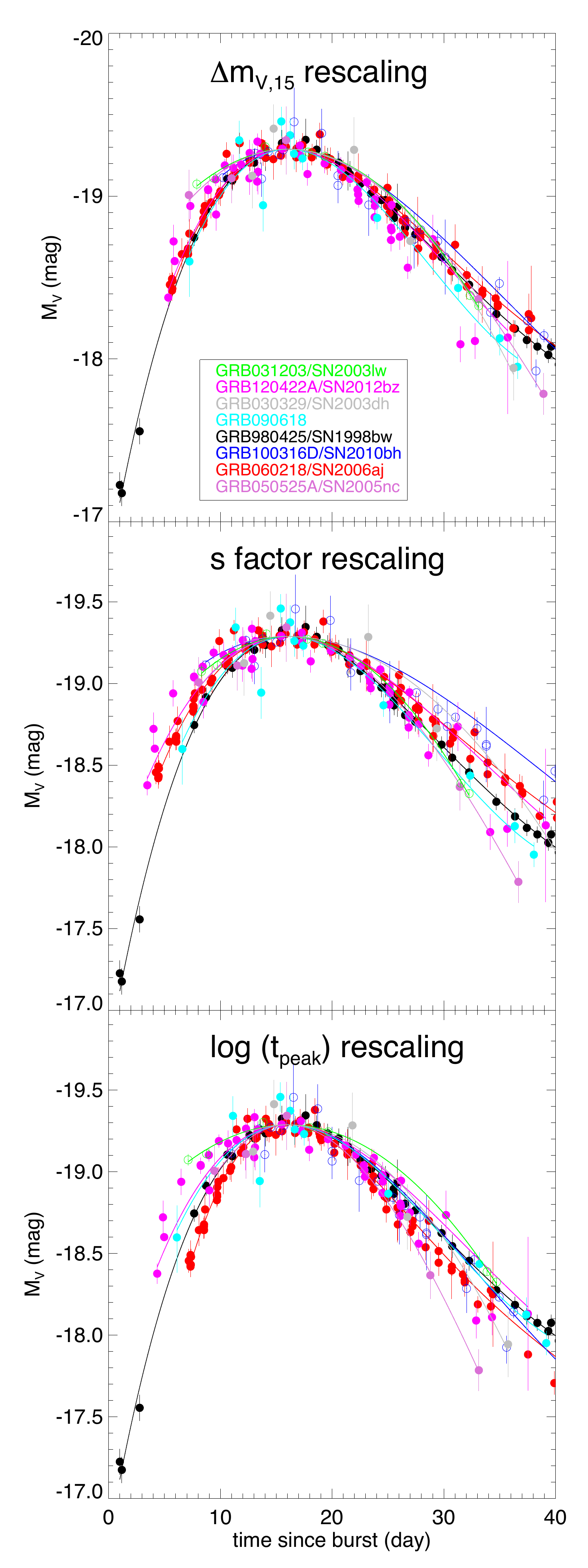}
\caption{    %${\Delta  m_{15}}$   rescaling (up) and   ${s}$ factor rescaling (down) of light curves of the selected systems in rest frame V band.  
Different rescaling results of the the light curves  of the selected systems in rest frame V band.  
Here    $M_{\rm V,peak}$ has been normalized   to that of SN 1998bw.   }
  \label{lightcurveall}
\end{figure}
%%%%%%%%%%%%%%%%%%%%%%%%%%%%%%%%%%%%%%%%%%%%%%%%

%(\emph {down}
 %%%%%%%%%%%%%%%%%%%%%%%%%%%%%%%%%%%%%%%%%%%%%%%%
%   \begin{figure*}
% \centering
%\includegraphics [width = 8.8 cm] {relationplot.pdf}
%\caption{   The peak-luminosity and decline rate relation for GRB-SNe. The big blue points represent for the GRB-SNe. The peak magnitude ($M_{\rm V,peak}$) and the decline rate ($\Delta m_{\rm V,15}$) (here $\alpha = 15$ day) are constrained with the method and the procedures discussed above. The small grey points in the background are for SNe Ia  \citep{hicken_2009}. 
%}
%  \label{relationplot}
%\end{figure*}
%%%%%%%%%%%%%%%%%%%%%%%%%%%%%%%%%%%%%%%%%%%%%%%%

 %%%%%%%%%%%%%%%%%%%%%%%%%%%%%%%%%%%%%%%%%%%%%%%%
   \begin{figure}
 %\centering
\includegraphics [width = 8.8 cm]  {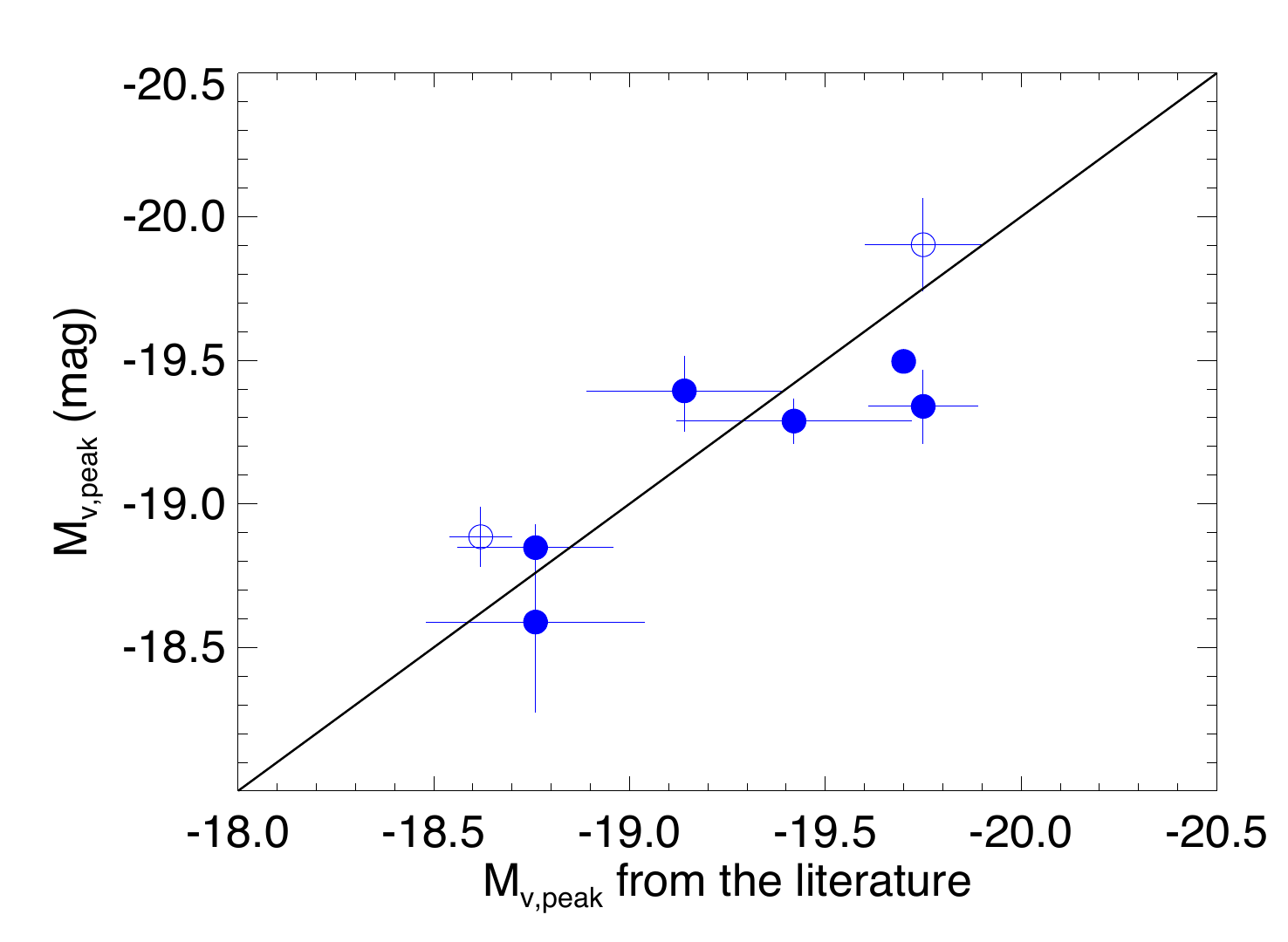}
\caption{   Comparison of the peak magnitudes determined in this paper ({\it y-axis}) and the  values from \cite{malesani_2003lw_one, cano_2010bh, cano_060729, canoz_2013, schulze_2012bz} ({\it x-axis}).   The illustrative straight line in black is: $y = x$.  Three outliers are: GRB 090618,  XRF 100316D/SN 2010bh and GRB 120422A/SN 2012bz.
 }
  \label{peakcp}
\end{figure}
%%%%%%%%%%%%%%%%%%%%%%%%%%%%%%%%%%%%%%%%%%%%%%%%

 %%%%%%%%%%%%%%%%%%%%%%%%%%%%%%%%%%%%%%%%%%%%%%%%
   \begin{figure}
 %\centering
\includegraphics [width = 8.8 cm]    {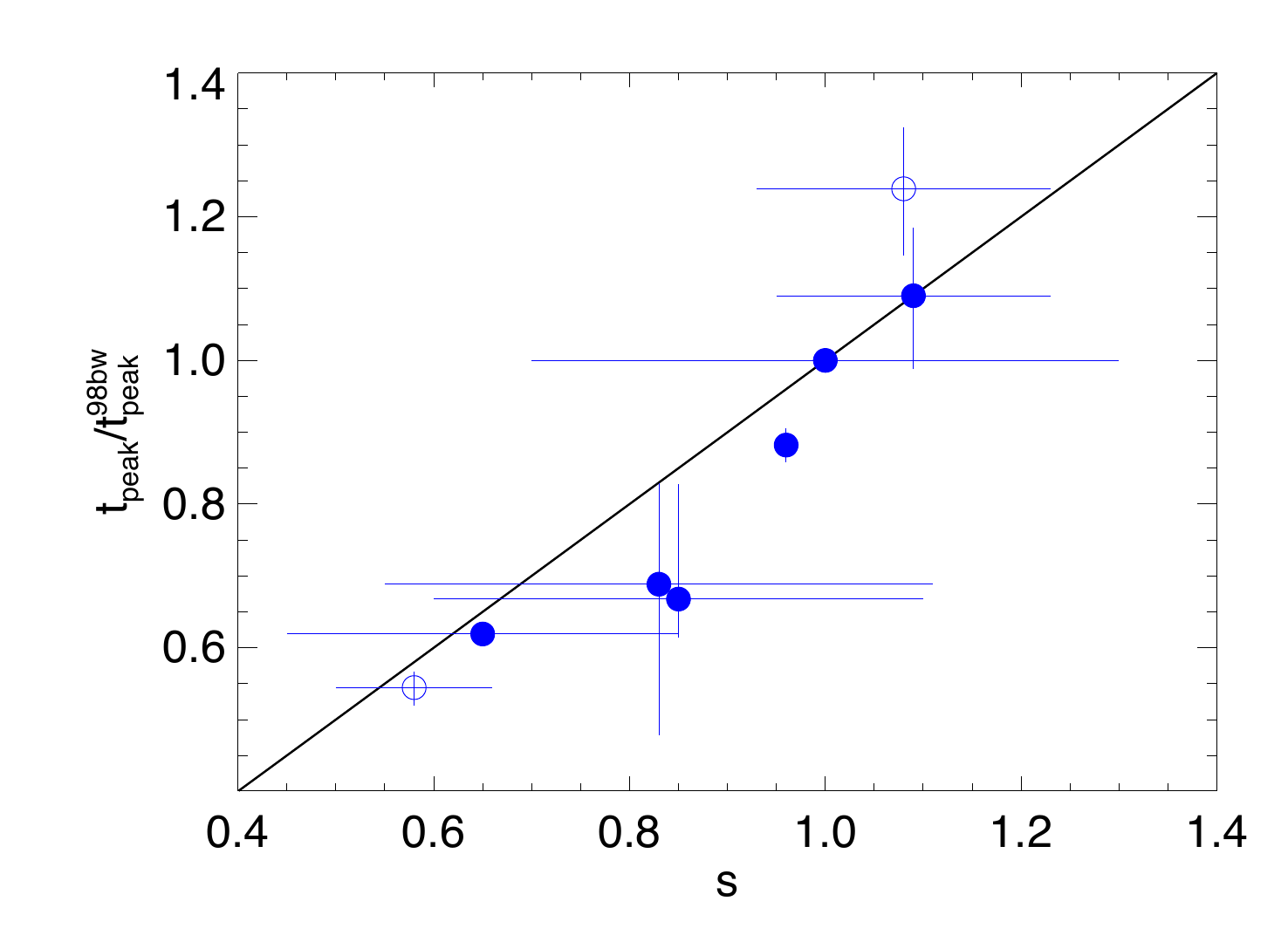}
\caption{   Comparison of peak time  $t_{\rm peak}$  obtained in this paper (in Table \ref{property}) and the stretch factor  $s$  from \cite{canoz_2013}.  Here $t_{\rm peak}^{\rm 98bw} = 16.09^{+0.17}_{-0.18}$ days. The illustrative straight line in black is $y = x$.    }
  \label{stretchtpeak}
\end{figure}
%%%%%%%%%%%%%%%%%%%%%%%%%%%%%%%%%%%%%%%%%%%%%%%%

 %%%%%%%%%%%%%%%%%%%%%%%%%%%%%%%%%%%%%%%%%%%%%%%%

The peak magnitudes, the decline rate in 15 days, the time of peak (see section \ref{subsec:tpeak}) of  the eight GRB-SNe are listed in Table \ref{property}.  %The table shows that GRB-SNe  and  SNe Ia are about equally bright at peak. 
  For     SNe Ia, there is a relation between the intrinsic peak magnitude $M_{\rm V,peak}$ and decline rate  $\Delta m_{\rm B,15}$  \citep{phillips_1993, phillip_1999}. 
  In addition,  % to  rescaling the light curves of  SN 1998bw with   the     $s$ factor,  
  $\Delta m_{\rm V,15}$ is used   in this paper to check if the light curves of SN 1998bw can be used as light curve templates and if there is a better way to do the rescaling other than using   the     $s$ factor.  

\subsection{Luminosity-decline rate    relation}
\label{subsec:ldrelation}

%\subsubsection{$\Delta m_{\rm V,\alpha}$}
\subsubsection{      $M_{\rm V,peak} = f(\Delta m_{\rm V,15})$       }
\label{sub:deltam}

Though   the physical progenitors and   explosion mechanisms for SNe Ia and GRB-SNe are   different  \citep{hillebrandt_2000, smartt_2009},  their light curves show   similar    luminosity-decline rate relations.  
The peak magnitude and the decline rate are resimulated 10 000 times each. The widths of the distribution of the resimulated data are $ 1  \sigma$. We linearly fit each set of the resimulated data and get two distributions of the fitting parameters. The median values and $\pm 1  \sigma$ values on two sides of the median values in these two distributions are treated as the best fitting parameters and the $\pm 1  \sigma$ uncertainties.   
The luminosity-decline rate    relation for GRB-SNe is 
    \begin{equation} 
%M_{\rm V,peak} = (1.50  \pm 0.18)  \Delta m_{\rm V,15} - (20.42 \pm 0.17).   
%M_{\rm V,peak} = (1.69  \pm 0.23)  \Delta m_{\rm V,15} - (20.63 \pm 0.22).   
%M_{\rm V,peak} = (1.38  \pm 0.22)  \Delta m_{\rm V,15} - (20.33 \pm 0.21). 
 % M_{\rm V,peak} = (1.63  \pm 0.19)  \Delta m_{\rm V,15} - (20.62 \pm 0.16). 
 % M_{\rm V,peak} = (1.76  \pm 0.20)  \Delta m_{\rm V,15} - (20.74 \pm 0.17). 
   %M_{\rm V,peak} = (1.78  \pm 0.75)  \Delta m_{\rm V,15} - (20.74 \pm 0.68). 
   M_{\rm V,peak} =  1.59^{+0.28}_{-0.24}   \Delta m_{\rm V,15} -  20.61^{+0.19}_{-0.22},  
 \label{peakdeclinelow}
\end{equation} 
with $\chi^2 = 5.2$ (6 dof).     
    Figure \ref{relationplotall} shows the luminosity-decline rate    relation with  
 $\alpha = $ 5, 10, and 15 days.       
 Systems GRB 031203/SN 2003lw and  GRB 100316D/SN 2010bh have uncertain extinction, so they are plotted as open symbols. Unless mentioned otherwise, these two systems are discerned in the same way in the following figures.   
  Some systems lack  data to constrain light curves at large  times,       i.e.,       $\alpha > 15$ days, so we cannot get $\Delta m_{\rm V, > 15}$ for these systems.  
This relation shows that 
(1) the peak magnitudes span a   small range;  (2)  the trend of the relation       is the same as for SNe Ia, i.e., brighter systems decline slower.    
%The rms dispersion in the relation for eight  GRB-SNe   is small ($\sigma \approx 0.37$ mag),  
Though there could be significant selection effects, in that we only have good data for bright systems.

%In this paper, besides $\alpha = 15$, we also  study the decline rate at 5 and 10 days. 
 %In this paper, we  study the time interval $\alpha \in [3, 15]$ days.     %At $\alpha = 15$, the relation has the smallest dispersion. 

 %The $\chi^2$ of the  luminosity-decline rate    relation  for the corresponding  time intervals are shown in Figure \ref{chisquare}.  Among the time intervals  from 3 to 15 days, at $\alpha = 15$ days, the peak magnitude and the decline rate have the strongest     correlation.  
%

\subsubsection{Correlation coefficients and significance}
 \label{sub:sig}

With the standard Monte Carlo method,   the correlation coefficients: Pearson's,  Kendal's $\tau$ and  Spearman's rank,        are calculated to statistically measure  the strength of  
 the correlation between the peak magnitudes and the decline rates.   
   As in section   \ref{sub:deltam},  the peak magnitude $M_{\rm V,peak}$  and the decline rate  $\Delta m_{\rm V, 15}$  are resimulated 10 000 times each, where $  1  \sigma$ are the   widths of the distribution of the resimulated data.   
%where the centre of peaks in each dimension are the parameter esti- mates, and the corresponding 1? errors are the width of the distributions. 
  For each set of resimulated data, we calculate three correlation coefficients.

Pearson's correlation coefficient (Pearson's $r$) measures the linear correlation   between two variables. The result  $r \in [-1, 1]$,  where 1 ($-1$)  is total positive (negative) correlation, and 0 is no correlation. %Different ranges of  $r$   give  different significance levels. 
 When  $|r| > 0.7$ (or 0.8 by different suggestion), the correlation is  described as `very strong'. If bin size is set to 0.01 (same in the following),     $r =  0.905$      has  the highest frequency and 
  this shows the correlation is significant at 0.01 level.     
  This means we expect to get the result occurring    by chance   once every 100 times.  
  The result  indicates a   significant correlation between $M_{\rm V,peak}$  and $\Delta m_{\rm V, 15}$.   
  There are [93\%, 87\%, 63\%, 9\%] of Pearson's correlation coefficients lie at [0.1, 0.05, 0.01, 0.001] significance levels.

Kendal's rank   correlation coefficient       (Kendal's $\tau$)  measures the strength of  the monotonic relationship between variables. 
  The result   $\tau \in [-1,1]$, where    $1/-1$  imply  the perfect agreement/disagreement  between two rankings and 0 means the ranking is totally independent.   In this paper,     $\tau =  0.715$  has the highest frequency and the corresponding significance level is 0.01.   
    There are [91\%, 79\%, 43\%, 0.3\%] of Kendal's rank correlation coefficients lie at [0.1, 0.05, 0.01, 0.001] significance levels.

Spearman's rank correlation coefficient    (Spearman's $\rho$) tests the   dependence   between two variables.  The result $\rho \in [-1,1]$,  where  $-1$ or $+1$ appears when the relation of the variables  can be perfectly described with a monotonic function.  When   $|\rho| \geq 0.6$, the correlation is  described as `strong'.   %(0.8), the correlation is  described as `strong' (`very strong').  
In this paper,  $\rho =  0.785$   has the highest frequency and the corresponding significance level is 0.025.  There are [95\%, 87\%, 30\%, 2\%] of Spearman's rank correlation coefficients lie at [0.1, 0.05, 0.01, 0.001] significance levels.

\begin{table}[h!tbp]
\begin{center}
\caption{The most frequent values and the percentage of the correlation coefficients %: Kendal's $\tau$, Spearman's rank, and   Pearson's  
at  different significance levels.  Here we set the bin size equal to 0.01.    }
 \label{tab:coefficient}
 %\begin{tabular}{c|cccc|c} 
 \begin{tabular}{cccccc} 

%Kendal's $\tau$  &  Spearman's rank  &  and   Pearson's     
%\begin{tabular}    {cccccccc@{}}
\hline       
\hline   
 % GRB/XRF/SN &    $z$ &  $M_{\rm V,peak}$      &  $\Delta m_{\rm V,15}$  & $t_{\rm peak}$ &  $ E(B-V)_{\rm MW}$    & $E(B-V)_{\rm host}$        \\ %\tnote{1} \\ 

coefficient &  \multicolumn{4}{ c  } {significance level$^a$}  &  most frequent value\\
                     &        0.1 & 0.05 &  0.01  & 0.001   &  \\ %\tnote{1} \\ 

      \hline
  \multicolumn{6}{ c } {luminosity-decline rate relation} \\
  \hline
Pearson's   & 93\%& 87\%& 63\%& 9\% & 0.905\\
Kendal's $\tau$  & 91\%& 79\%& 43\% &  0.3\%   & 0.715   \\
Spearman's rank  & 95\%& 87\%& 30\%& 2\% &0.785 \\
  \hline
% \hline
      
 \multicolumn{6}{ c } {$k-s$ relation} \\
  \hline
Pearson's   & 99\% &  97\% &        14\% &  $\sim$ 0\% & 0.795\\
Kendal's $\tau$  & 60\% &  27\% &  1.5\% & $\sim$ 0\%     & 0.505   \\
Spearman's rank  & 99\% &    57\% & 3\% &  $\sim$ 0\%  &0.595 \\
   \hline
% \hline
                                                                        
\end{tabular}
\end{center}
{\footnotesize
 \noindent
  $^a$: The probability of  accidentally getting the result, e.g., 0.05 represents the result happens by chance   once every 20 times.   
}
\end{table}

The distributions of the correlation coefficients  of the luminosity-decline rate relation  
are plotted in the upper panel in Figure \ref{coefficient}. The  most frequencies and the percentage of the correlation coefficients at different significance levels    are listed in Table \ref{tab:coefficient}.   The statistical correlation coefficients show that the luminosity and the decline rate   of GRB-SNe    are significantly correlated.

 %%%%%%%%%%%%%%%%%%%%%%%%%%%%%%%%%%%%%%%%%%%%%%%%%

%We conclude that the  peak magnitudes and the decline rates are strongly correlated. 

%\subsubsection{ $\alpha $ at other days}
\subsection{Time since burst}
\label{sub:tsb}
\subsubsection{Peak time}
\label{subsec:tpeak}
The peak time $t_{\rm peak}$ is defined as the time when the light curve of a GRB-SN  reaches its peak brightness    relative to the time of the GRB        in the rest frame. 
%  The time when the light curve of a GRB-SN  reaches its peak brightness since the burst in the rest frame  is defined as the peak time $t_{\rm peak}$. 
  The peak times for  the eight systems  are  listed in Table \ref{property}.    
     With the same procedure as  in  section  \ref{sub:deltam},     
 the best fit  to the relation between   $\log t_{\rm peak}$ and $M_{\rm V,peak}$    is
      \begin{equation} 
  M_{\rm V,peak} =  - 2.52^{+0.16}_{-0.15}   \log t_{\rm peak}   - 16.41^{+0.16}_{-0.18}. 
  \label{peaklogt}
\end{equation} 
 When combining the two parameters $\log t_{\rm peak}$ and $\Delta m_{\rm V,15}$,   regression fit    to  $M_{\rm V,peak}$ can be expressed as
        \begin{equation} 
 M_{\rm V,peak}=   1.46^{+0.73}_{-0.88}   \Delta m_{\rm V,15} -  0.29^{+1.01}_{-1.41}  \log t_{\rm peak}   -  20.19^{+2.34}_{-1.74}. 
  \label{mvm15logt}
\end{equation} 
The relation between $\log t_{\rm peak}$ and $\Delta m_{\rm V,15}$ can be expressed as 
       \begin{equation} 
 \log t_{\rm peak} =  - 0.51^{+0.13}_{-0.11}   \Delta m_{\rm V,15}   +  1.56^{+0.09}_{-0.10}. 
  \label{m15logt}
  \end{equation} 
Figure~\ref{logtpeak}    shows the relations between $\log t_{\rm peak}$, $\Delta m_{\rm V,15}$ and $M_{\rm V,peak}$.  The upper panel shows  that there is a dependency  between  the peak time  $t_{\rm peak}$   and   the  peak magnitude $M_{\rm V,peak}$.        There is a trend that a GRB-SN  with    smaller peak time  has   fainter peak luminosity, i.e.,   $M_{\rm V,peak}$  decreases as  $t_{\rm peak}$ increases. In general,  brighter  GRB-SNe   evolve more slowly.  
  Compared to Figure \ref{relationplotall},    $t_{\rm peak}$ is less strongly  correlated with   $M_{\rm V,peak}$ than  $\Delta m_{\rm V,15}$.  The middle panel shows a multiple linear regression fit  to  $M_{\rm V,peak}$ with $\log t_{\rm peak}$ and  $\Delta m_{\rm V,15}$. %The multiple linear regression fitting function is in Eq. \ref{mvm15logt}.   
The bottom panel of Figure~\ref{logtpeak} shows  a `fundamental plane' of GRB-SNe with  peak time   $t_{\rm peak}$  and   decline rate  $\Delta m_{\rm V,15}$. Constant absolute peak magnitudes are also indicated by dotted lines.

 %{sub:deltam} {sub:tsb} {sub:sfactor}

\subsubsection{   $k-s$ relation }
 
\label{sub:sfactor}

 Besides the peak magnitude  $M_{\rm V,peak}$  and  the decline rate $\Delta m_{\rm V,\alpha}$, another way to describe the light curve is  through the luminosity factor  $k$  and the   stretch factor  $s$.  These two factors stand for  the relative peak    ($k$)  and  width   ($s$)  of the light curves compared to SN 1998bw \citep{cano_2010bh}:
     \begin{equation} 
f (t) = k \times f^{\rm 98bw} (t/s), 
  \label{tsfactor}
\end{equation}  
 Here $f(t)$ is the flux of a SN,  and $f^{\rm 98bw}(t)$ is the flux of SN 1998bw.  The factor $s$  equals to $t_{\rm peak} /t_{\rm peak}^{\rm 98bw}$,  with $   t_{\rm peak}^{\rm 98bw} $ representing the peak time of SN 1998bw.  %in   \ref{subsec:lcall}. 
  With the same procedure as  in  section \ref{sub:deltam},  the $s$ and $k$ factors are correlated as 
     \begin{equation} 
%k = (1.18 \pm 0.64) \cdot s - ( 0.01 \pm 0.48).  
k = 1.25^{+0.12}_{-0.12}  \cdot s - 0.05^{+0.09}_{-0.09}, 
  \label{eq:ksfactor}
\end{equation}  
 with $\chi^2 = 8.2$ (6 dof).  
 This relation is named as  $k-s$ relation.    Figure \ref{ksfactor} shows the  correlations   between       $s$,    $k$, the decline rate $\Delta m_{\rm V,15}$ and the  peak magnitude $M_{\rm V,peak}$. 
 %The relations between  the stretch factor  and the peak luminosity  are shown in Figure \ref{speak}.  % and \ref{stretchcp}

 With the   procedure discussed in section \ref{sub:sig},   
the distributions of the correlation coefficients  of the $k-s$  relation  
are plotted in the lower panel in Figure \ref{coefficient}. The  most frequencies and the percentage of the correlation coefficients at different  significance levels  are listed in Table \ref{tab:coefficient}.   %
       Comparing the results in Figures   \ref{relationplotall},    \ref{logtpeak},   \ref{ksfactor} and Table \ref{tab:coefficient},  %Figures \ref{relationplotall},  \ref{logtpeak} and \ref{speak} and Eqs.  \ref{} and, 
    we conclude   that  1)  the correlation between $\Delta m_{\rm V,15}$ and  $M_{\rm V, peak}$   is stronger   than the one between  factors $k$ and $s$.  
    2) $\Delta m_{\rm V,15}$  is stronger correlated with $M_{\rm V, peak}$   than $t_{\rm peak}$ and  the $s$ factor.

\subsection{Rescaling of light curves}
\label{subsec:lcall}

  To test if the light curve of SN 1998bw  can be used as a template for other GRB-SN light curves, we rescale
  the light curves in three ways.  In all cases,    $M_{\rm V,peak}$ has been normalized relative to SN 1998bw.
 \begin{description}
  \item[   $\bm{\Delta  m_{15}}$  \textbf{rescaling}]  The light curves are rescaled around $t_{\rm peak}$.    The   time of the light curve is calculated as $t' =  (t - t_{\rm peak})  \times   \Delta m_{\rm V,15} / \Delta  m_{15}^{\rm 98bw} + t_{\rm peak}^{\rm 98bw}$, with $ \Delta    m_{15}^{\rm 98bw}$ and $ t_{\rm peak}^{\rm 98bw}$ representing the decline rate and the peak time of SN 1998bw.  

    \item[\textbf{$\bm{s}$ factor rescaling}]  The light curves are rescaled  as $t' = t /  s$,  with $s$ being the stretch factor. % (Eq. \ref{tsfactor}).  %t_{\rm peak}^{\rm 98bw} /t_{\rm peak}$.     %The term $t_{\rm peak}^{\rm 98bw} /t_{\rm peak}$ denotes the  $s$ factor discussed in section \ref{sub:sfactor}. 

 % \item[   $\bm{ t_{\rm peak}}$  \textbf{rescaling}]  The light curves are rescaled around $t_{\rm peak}$.    The time has been calculated as $t' =  (t - t_{\rm peak})  / s$, with $s$ being the stretch factor (Eq. \ref{tsfactor}).   

  \item[   $\bm{ \log (t_{\rm peak})}$  \textbf{rescaling}]  The light curves are rescaled around $t_{\rm peak}$.    The time is calculated as $t' =  (t - t_{\rm peak})  \times  \log  (t_{\rm peak}^{\rm 98bw}) / \log ( t_{\rm peak})  + t_{\rm peak}^{\rm 98bw}$. %, with   $ t_{\rm peak}^{\rm 98bw}$ representing   the peak time of SN 1998bw.  

    \end{description}

 A collection of rescaled light curves for the selected systems are shown in 
 Figure~\ref{lightcurveall}.    
 The fitting curves are    the rest frame V band light curves obtained in     section  \ref{sec:sys}.  
 The data points are for illustration and are from the bands closest to the rest frame V band.  
The figure  shows   that a rescaled SN 1998bw light curve is a reasonable template for other GRB-SN light curve. 
$\Delta  m_{15}$  rescaling appears  superior to the other approaches.   %If rescaling with stretch factor $s$, there is not much difference at rescaling since the time of burst or rescaling around the peak time. 
 If    values of $\Delta  m_{15}$ are not available,    $ \log (t_{\rm peak})$  rescaling is an alternative to the commonly used  $s$ factor rescaling.

\subsection{Discussion}
\label{sub:dis}
   
 We compare the values of the peak magnitudes to other studies  %\citep{malesani_2003lw_one, cano_2010bh, cano_060729, canoz_2013, schulze_2012bz, giorgos_2014}.  
  \citep{malesani_2003lw_one, cano_2010bh, cano_060729, canoz_2013, schulze_2012bz}.    %, giorgos_2014
  The result  is shown in Figure~\ref{peakcp}. There are three obvious outliers:     GRB 090618, SN 2010bh and SN 2012bz. These   systems are estimated to have fainter peak magnitudes in this paper.   It is difficult to trace the 
     exact  causes of  the differences in the peak magnitudes.  We follow the procedure from the literature, and compare it to our results.  %The reasons may come from but not limited to the ones listed in the following discussion. 

 There is no independent third-part study for the   peak magnitude of GRB 090618.  The peak magnitude is estimated to be $M_{\rm V, peak} =   -19.34^{+0.13}_{-0.13} \,  (-19.75_{-0.14}^{+0.14})$ mag from this paper \citep{cano_060729}.   The  reasons cause the difference are: 
 1) With different cosmological parameters, the distance modulus is different. We adopt   distance modulus   $\mu =$  42.53    with the cosmological parameters   \{$\Omega_{M}, \Omega_{\Lambda}, h$\} = \{0.315, 0.685, 0.673\}, while $ \mu = $  42.45 from  \cite{cano_060729}.  
 2) The subtraction of the afterglow   may be another reason for the discrepancy.   This may cause  $\sim 0.15$ mag difference around the peak.  The observed peak magnitude is $i = 22.33$ mag. After the host  and the  afterglow subtraction with $i_{\rm host} = 23.22 \pm 0.06$ mag,  the peak magnitude becomes $i = 22.96$ mag and $i = 23.13$ mag, respectively. From \cite{cano_060729} the apparent peak magnitude is $i = 23.00$ mag. 
 3) The polynomial fitting may bring $\sim 0.08$ mag difference. The fitted (observed) apparent peak magnitude is   $i = 23.21$ (23.13) mag. Figure \ref{fig:grb050525} shows the observed data. Around peak, the data are noisy.    At $t =$   16.69 and  17.60 days, the magnitudes are $\sim 0.2$  and 0.04 mag fainter than   the one at $t=$ 14.79.     
%Many effects may cause the noisy data, e.g., observed condition may change from day to day.  
 Instead of using a single datum, in this paper,  polynomial functions   are fitted, especially around the peak. %to smooth the   data. 
 4)   The Galactic extinction is     different. In this paper, we use  $R_V = 3.1$ as well as the re-calibration results  of DIRBE/IRAS  dust map  from  \cite{schlafly_dustmap}.    %\cite{cano_060729} adopts $R_V = 2.93$ \citep{pei_rv}.  
 This may bring $\sim 0.15$ mag difference.  
% 5)   The term $2.5 \log (1+z) = 0.47$ is added in K-correction  (Eq. \ref{magthres}).  

 %%$-19.50^{+0.04}_{-0.05}$   $-19.31^{+0.13}_{-0.13}$

 For XRF 100316D/SN 2010bh, the peak magnitude is constrained to be  $M_{\rm V, peak} =  -18.89^{+0.10}_{-0.10}$/$-18.62 \pm 0.08$ mag  in this paper/\cite{cano_2010bh}. This is a system with uncertain host extinction and peak magnitude. The reasons are:
 1) The values in  \cite{cano_2010bh} are inconsistent. In Table 2 \citep{cano_2010bh},  the apparent peak magnitude in the V band is 19.47 mag after the Galactic extinction correction. So if     $R_V = 3.1$ and K correction $\Delta k = 0.09$ are used in his calculation,   the peak magnitude should be $M_{\rm V, peak} =  m_V - \mu - A_{\rm V, host}  - \Delta K = 19.47 - 37.08 -  0.18 \cdot 3.1  - 0.09 =  -18.26$ mag instead of $-19.62$  mag, which is listed   
 In Table 4   \citep{cano_2010bh}.  The photometric data from  \cite{cano_2010bh} is consistent with  the result from \cite{bufano_2010bh}, so we guess the foreground extinction is   subtracted twice in the calculation, which is consistent  with the  statement of the   captions of Table 2 and 4.   The result also shows that a larger host extinction is expected, otherwise the spectrum is very red (as stated in section  \ref{sub:2010}). 
 2) The host extinction estimated in the literature are different. We adopt a large extinction with $E(B-V)_{\rm host} = 0.39 \pm 0.03$ mag from  \cite{olivares_e_2010bh}.  The value of $E(B-V)_{\rm host} = 0.18 \pm 0.08$ mag is estimated in  \cite{cano_2010bh}. This may cause $\sim 0.64$ mag difference.
  3)  The distance moduli  are different. In this paper, $\mu = 37.20$ while \cite{cano_2010bh} adopts $\mu = 37.08$.  This causes about 0.12 mag difference. 
4) The K correction in  \citep{cano_2010bh} may bring about 0.09 mag difference.

For GRB 120422A/SN 2012bz, the peak magnitude is estimated to be   $M_{\rm V, peak} =  -19.50^{+0.03}_{-0.03}$/$-19.7$ mag    in this paper/\cite{schulze_2012bz}, while using the same cosmological parameters \citep{planck_cosmopara}, \cite{giorgos_2014} estimates the peak magnitude to be $-19.63$ mag.  
The reasons of the discrepancies may be as follows: 
1)   The Galactic extinction  may be corrected twice in calculating the absolute peak magnitude \citep{schulze_2012bz, giorgos_2014}. So the magnitude should be about 0.09 mag fainter.  
2) In addition, $R_V$  instead of $R_I$ may be multiplied in calculating the Galactic extinction.   This may in further bring about $\sim 0.05$ mag difference.  
3) As discussed above, the noisy data around the peak    may bring about 0.03 mag difference.  
%4) The convert from $V$ to $V_{\rm AB}$ by 0.02 mag may be another reason.  

 %In the  K-correction (Eq. \ref{magthres}), an extra term $2.5 \log (1+z) = 0.27$ mag  is added to the peak magnitude. This may be the biggest contribution to the difference of the estimate of the peak magnitude. The difference may also come  from different power-law functions fitted to the afterglow.  The slope is fixed by \cite{schulze_2012bz} with $\beta_2 = -2$, based on SN modeling, while we fix the slopes of afterglow in the r'  and i' bands to be $\beta_2 = -1.48 \pm 0.4$, based on the X-ray light curve.  %This may also the reason why the stretch factor $s$ are different in Figure \ref{stretchtpeak},  compared to our study and \cite{canoz_2013}.  %, considering SN 2012bz has $z = 0.283$, therefore the peak magnitude of SN 2012bz is 0.27 mag fainter than \cite{schulze_2012bz}  in K correction step. 

  Figure \ref{stretchtpeak} shows the comparison of peak time in this paper (in Table \ref{property}) and the stretch factor  $s$ from  \cite{canoz_2013}.  There  is  no information on $s_V$ for  systems SN 2003dh, SN 2003lw,  GRB 050525A  and GRB 090618,  so $s_R$ is used instead. %The result shows that except for SN 2012bz,  the $s$ factor ($t_{\rm peak}/t_{\rm peak}^{\rm 98bw}$) constrained in this paper is consistent with the results from  \cite{cano_2010bh}. For SN 2012bz, the difference of $s$ factors may come from different afterglow fitting slopes.  

%===============================================================
% Conclusions
%===============================================================

\section{Conclusions}
\label{sec:summary}
 
We   developed a method for obtaining the light curves in the rest frame V band from the observational data.    A standard Monte Carlo method was used for error estimation.    Afterglow and host brightness were subtracted.  We used the   DIRBE/IRAS  dust map   and the correction coefficients        to correct  the    foreground extinction.  
The host extinction was corrected.    
We used a multi-band K-correction to correct the light curves from the observed bands into the rest frame V band. Alternatively, SN 1998bw peak SED and decline rate templates  were used when a multi-band K-correction is not feasible.  Polynomial functions were fitted to obtain the light curves.

 %DIRBE/IRAS   dust map \citep{schlegel_dustmap} and the correction factors  \citep{schlegel_dustmap}

Based on this method we   obtained the peak magnitudes and the decline rates  for eight GRB-SN systems in classes  $A$, $B$, and $C$.
We   discovered a     relation  between the peak magnitude and the decline rate.    This luminosity-decline rate    relation was tested   with the decline day $\alpha$ at 5, 10 and 15 days.  
 The strength of the relationship between the peak magnitude and the decline rate was statistically measured by three correlation coefficients  and the significance levels were discussed.   %from 3 to  15 days.    %At $\alpha = 15$ days, the correlation is strongest      with $ \chi^2 = 1.21$.  
%The luminosity-decline rate    relation  has smaller $\chi^2$ than the $k-s$ relation.   
There   is a dependency between the peak magnitude and the peak time. The larger the peak time, the brighter the SN is.  We   found that the light curve of SN 1998bw can be used as a representative template. In addition, rescaling   around the peak time with   $\Delta m_{\rm V,15}$ is  better than  rescaling with  peak time $\log t_{\rm peak}$ or stretch factor $s$.     We also compared the peak magnitudes  and the decline rates constrained from this work to the results from other studies.

SNe Ia and GRB-SNe have completely different progenitors. Nevertheless, the light curves have   similar peak magnitudes and decline rates. This phenomenon may potentially help us    shed light on progenitor models of GRBs.    % If more data beyond $\alpha = 15$ days are collected, we may further test if the  luminosity-decline rate   relation is most strongly correlated at $\alpha = 15$ days.  
 As SNe Ia are widely used as standard candles to measure cosmological distances, it is possible that GRB-SNe may also turn out to be useful high-redshift standard candles. 
In particular, the prospects of studying dark energy through $w(z)$ with GRB-SNe using the James Webb Space Telescope (JWST) is intriguing.
%Since the connection between GRB and SN is attracting more and more attention,  as well as the associated    GRBs have     extraordinary brightness,        GRB-SNe will become a very powerful tool in  cosmology. 

%===============================================================
% Acknowledgements
%===============================================================

\begin{acknowledgements}

 \label{Acknowledgments}

 We thank  Enrico Ramirez-Ruiz, Tamara Davis    Giorgos Leloudas and  Radek Wojtak  
  for their many helpful discussions and comments on the paper.   
 We thank Teddy Frederiksen, Darach Watson, Daniele Malesani  and Dong Xu for     discussions on GRB and SNe.    
 The Dark Cosmology Centre is funded by the Danish National Research Foundation.

\end{acknowledgements}

%===============================================================
% Bibliography
%===============================================================

\bibliographystyle{aa}
\bibliography{ref}

\end{document}